\documentclass[fleqn,usenatbib]{mnras}

\usepackage{newtxtext,newtxmath}

\usepackage[T1]{fontenc}
\usepackage{ae,aecompl}

\usepackage{graphicx}	
\usepackage{amsmath}	
\usepackage{mathtools}
\usepackage[normalem]{ulem}

\usepackage{mathtools}

\usepackage{subcaption}
\newcommand{\thetalog}{$(\theta, \log_{10}(R+1))$}
\newcommand{\Rpar}{$\mathcal{R}$}
\newcommand{\CR}{$R_{CR}$}
\newcommand{\Om}{$\Omega_{Bar}$}
\newcommand{\QBar}{$Q_{Bar}$}

\def\Col#1{Col. (#1):}
\def\code#1{\texttt{#1}}

\def\Pval#1#2{$(#1 \times 10 ^{#2})$}

\title[How the bar affect the spiral structure]{How the bar properties affect the induced spiral structure}

\author[L. Garma-Oehmichen et al.]{
L. Garma-Oehmichen,$^{1}$\thanks{E-mail: lgarma@astro.unam.mx}
L. Martinez-Medina,$^{1}$
H. Hern\'andez-Toledo,$^{1}$
and I. Puerari.$^{2}$
\\
$^{1}$Instituto de Astronom\'ia, Universidad Nacional Aut\'onoma de M\'exico, Apartado Postal 70-264, CDMX, 04510, M\'exico \\
$^{2}$Instituto Nacional de Astrof\'\i sica, Optica y Electr\'onica, Apdo. Postal 51 y 216, 72000 Puebla, Puebla, M\'exico}
\date{Accepted XXX. Received YYY; in original form ZZZ}

\pubyear{2021}

\begin{document}
\label{firstpage}
\pagerange{\pageref{firstpage}--\pageref{lastpage}}
\maketitle

\begin{abstract}
Stellar bars and spiral arms co-exist and co-evolve in most disc galaxies in the local Universe. However, the physical nature of this interaction remains a matter of debate. In this work, we present a set of numerical simulations based on isolated galactic models aimed to explore how the bar properties affect the induced spiral structure. We cover a large combination of bar properties, including the bar length, axial ratio, mass and rotation rate. We use three galactic models describing galaxies with rising, flat and declining rotation curves. We found that the pitch angle best correlates with the bar pattern speed and the spiral amplitude with the bar quadrupole moment. Our results suggest that galaxies with declining rotation curves are the most efficient forming grand design spiral structure, evidenced by spirals with larger amplitude and pitch angle. We also test the effects of the velocity ellipsoid in a subset of simulations. We found that as we increase the radial anisotropy, spirals increase their pitch angle but become less coherent with smaller amplitude.

\end{abstract}




\section{Introduction}
\label{sec:introduction}

Stellar bars inhabit a large fraction of disc galaxies in the local Universe, in a great variety of shapes, sizes, and environments. They promote the galaxy secular evolution by exchanging mass, energy and angular momentum with stars and gas across the disc \citep{Weinberg1985,Kormendy2004, Sellwood2014, Diaz-Garcia2016_Secular_Evolution}, transporting angular momentum from the inner bar resonances to those outside corotation \citep{LyndenBell1972, Tremaine1984b, Athanassoula2003} and redistributing stars in the disc by radial heating and radial migration \citep{Monari2016a, Martinez-Medina2017}. Bars can also induce gas inflow to the central regions via shock waves, forming gaseous structures like dust-lanes and rings \citep{Athanassoula1992, Hernquist1995, Martinent1997, Kim2012_gas, Sormani2018, Seo2019}. Some moving groups in the solar neighbourhood have their dynamical origin under the influence of the Galactic bar \citep{Perez-Villegas2017} and through its induced resonances shape the stellar velocities in the solar neighbourhood \citep{Fux2001}.

The importance of these phenomena depends on the morphological and dynamical characteristics of the bar. In late-type galaxies, bars tend to be smaller in relation to their discs \citep{MendezAbreu2012, Diaz2016_Bar_characteristics, Erwin2018}, more oblate shaped \citep{MendezAbreu2012, Diaz2016_Bar_characteristics} and prone to having an exponential density profile \citep{Elmegreen1985, Kim2015}. In contrast, bars in early-type galaxies tend to be larger, prolate shaped, and with a flat density profile. Such co-relations with the galaxy type have not been observed in measurements of the bar pattern speed (hereafter \Om{}), neither with the dimensionless rotation rate (hereafter parameter \Rpar{}) defined as the ratio between the bar length (hereafter $a$) and the corotation resonance (hereafter $R_{CR}$)  \citep{Aguerri2015, Guo2019, Cuomo2019, Garma-Oehmichen2020}. Nonetheless, it has been observed that the \Om{} seems to correlate with the galaxy luminosity and total stellar mass \citep{Garma-Oehmichen2020, Cuomo2020}. 

Spiral arms also play an important role in the secular evolution of galactic discs. By exchanging angular momentum, spirals churn stars and gas in the disc \citep{Sellwood2002}. There is a large body of evidence, theoretical and observational, that radial migration is an ubiquitous process in galaxies with spiral arms, central bar, or both \citep{Minchev2010, Vera-Ciro2014, Hayden2015, Loebman2016, Martinez-Medina2017, Daniel2018}. 
Spiral arms also induce dynamical heating in the disc, increasing the velocity dispersion with time \citep{Holmberg2009,Roskar2013,Martig2014} and modifying the velocity ellipsoid, an effect that depends on the nature and morphological properties of the spiral pattern \citep{Jenkins1990,Gerssen2012, Martinez-Medina2015}.  

In most barred galaxies, bars co-evolve with spiral arms, and in some cases, they could drive the spiral structure \citep{Salo2010}. This is especially apparent in two-armed grand design galaxies, where most spirals appear to be connected to the ends of the bar. How these structures interact remains a matter of debate. \cite{Kormendy1979} first suggested that strong bars or galaxy companions can lead to the formation of spiral density waves. There are different theoretical scenarios on how the pattern speed of both structures is related \citep[see e.g.][]{Dobbs2014}. If bars and spirals are strongly coupled, the pattern speed and amplitude of both structures should be strongly correlated. This strong coupling is expected from the spiral density theory \citep{Lin1964}, and the manifold theory where spirals are formed by stars in escaping orbits around the unstable Lagrangian points at the end of the bar \citep{Romero2006, Romero2007, Romero2015, Athanassoula2009}. Observational evidence of a strong coupling comes from the correlation between the torques and density amplitudes of both structures \citep{Buta2003, Block2004, Buta2005, Salo2010, Bittner2017, Diaz2019}. Another scenario proposes that the pattern speeds are related by a non-linear mechanism \citep{Tagger1987}, as suggested through indirect measurements of the bar pattern speed \citep{Font2017}. Finally, a third scenario where bar and spirals are decoupled structures, is supported by observations of the galaxy NGC 1365, where \citet{Speights2016} found results consistent with bar and spiral patterns being dynamically distinct features. The observational evidence to distinguish these scenarios is limited though, and hard to obtain due to the great uncertainties in estimating the pattern speed of both structures \citep{Garma-Oehmichen2020}.

The vast majority of N-body simulations show that spirals are transient or multi-arm features, that can be formed recurrently in time \citep{Sellwood2011, Grand2012, D'Onghia2013, Mata2019}. Numerical simulations with steady bar potentials have found that the gas particles settle in steady-state trailing spirals \citep{Sanders1976, Athanassoula1992, Wada1994, Rodriguez-Fernandez2008}. Collisionless  star particles can also produce prominent spirals and rings near the Outer Lindblad Resonance \citep{Schwarz1981, Bagley2009}. 

In this paper we use a set of numerical simulations based on galactic potential models with one million test particles to explore how the bar properties affect the response spiral arms. Our simulations cover a large bar parameter space, tailored to explore a large number of combinations of bar properties (size, shape, mass and pattern speed). We also test the effect of discs with different shapes of rotation curves and velocity ellipsoids. To characterise the response spiral arms, we identify the spiral particles using the density-based clustering algorithm \code{DBSCAN}. This let us obtain clean measurements of the spiral amplitude and pitch angle.

The paper is organised as follows. In Section \ref{sec:simulations} we describe the galactic potential models, and the space of parameters to explore. In Section \ref{sec:DBSCAN_spiral_properties} we describe how we use the algorithm \code{DBSCAN} to detect the spiral over-densities, and how we estimate the properties of the induced spiral arms. In Section \ref{sec:bar-spiral} we show how the bar parameters affect the spiral arms properties. In Section \ref{sec:rotation_influence} we explore the effects of different galactic models and the rotation curve shape. In Section \ref{sec:velocity_ellipsoid} we present the effects of the velocity ellipsoid. In Section \ref{sec:Bar_vs_Disc} we discuss the relative importance of the bar and disc properties in predicting the spiral properties. Finally, in Section \ref{sec:discussion} we discuss our results and present our conclusions.

\section{Simulations}
\label{sec:simulations}

\subsection{Galactic models}
\label{sec:galactic_models}

The mass distribution of a galaxy could play a significant role in the properties of the spiral arms. Using a sample of 94 galaxies \cite{Biviano1991}, observed that galaxies with steeper, rising rotation curves tend to host flocculent spirals, while galaxies with flat rotation curves have grand design spirals. \cite{Seigar2005, Seigar2006} found that the pitch angle strongly correlates with the rate of shear in the disc defined as

\begin{equation}
S=\frac{1}{2}\left( 1 - \frac{R}{V} \frac{dV}{dR} \right)    
\label{eq:Shear}
\end{equation}

They suggested that spirals in rising rotation curves galaxies have greater pitch angles compared to galaxies with flat or declining rotation curves. However, this relationship has been questioned by \cite{Kendall2015} and \cite{Yu2018}, who failed to observe such correlation (see also \cite{Diaz2019}).

To test the effects of different rotation curves, we used three galactic potential models with similar enclosed mass at $\sim$20 kpc, but different mass distributions. We will refer to these models as \textit{flat}, \textit{rising} and \textit{declining}, to figure out the shape of their rotation curve. The \textit{flat} model is the well-known \cite{Allen1991} potential comprising a Plummer bulge, a Miyamoto \- Nagai disc, and a spherical halo that reproduces the nearly flat Milky Way rotation curve. The density profiles of the bulge, disc and halo (referred with the sub-indexes B, D and H, respectively) are:

\begin{equation}
\rho_{B} (r) = \frac{3 b_B^2  M_B}{4 \pi \left(r^2 + b_B^2\right)^{5/2}}
\label{eq:bulge}
\end{equation}

\begin{align}
\rho_{D} (R, z) &= \frac{b_D^2  M_D}{4 \pi} \frac{\left(R^2 a_D + 3 \left(z^2 + b_D^2\right)^{1/2}\right)}{\left( R^2 + \left( a_D + \left(z^2 + b_D^2 \right)^{1/2}\right)^2\right)^{5/2}} \nonumber\\
& \times \frac{\left(a_D + \left(z^2 + b_D^2 \right)^{1/2} \right)^2} {\left(z^2 + b_D^2 \right)^{3/2}} 
\label{eq:disc}
\end{align}

\begin{equation}
\rho_{H} (r) = \frac{M_H}{4 \pi a_H r^2} \left(\frac{r}{a_H}\right)^{1.02} \left( \frac{2.02 + (r/a_H)^{1.02}}{(1+(r/a_H)^{1.02})^2}\right) 
\label{eq:halo}
\end{equation}

where $r$ is the spherical radius coordinate, $(R,z)$ are cylindrical coordinates, $M_B$, $M_D$ and $M_H$ are the masses, and $b_B$, $a_D$, $b_D$, $a_H$ are the characteristic scales of each component.

\begin{figure}
    \centering
    \includegraphics[width=\linewidth]{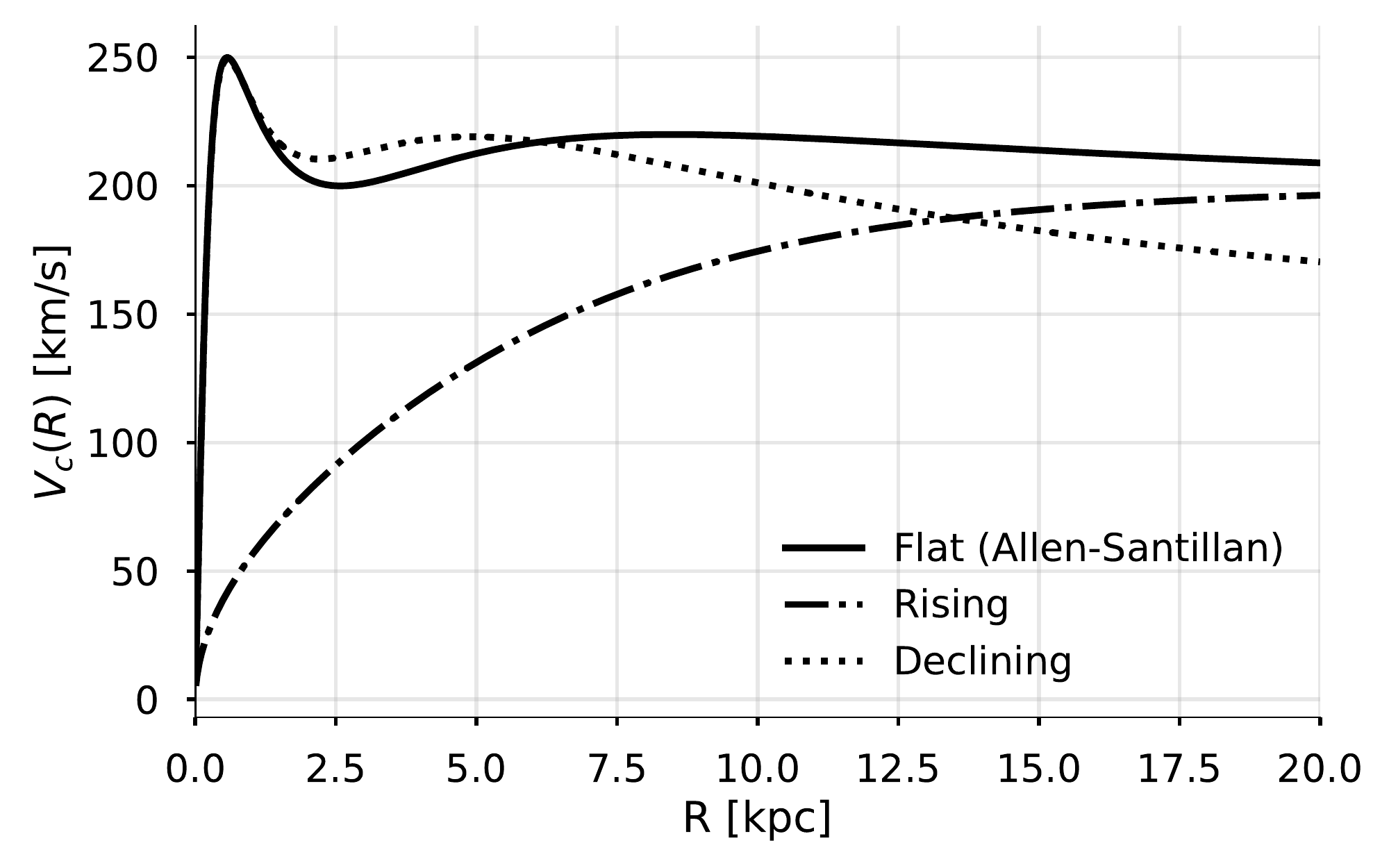}
    \caption{Rotation curves of the three axisymmetric galactic models used in this work.}
    \label{fig:rotation_curves}
\end{figure}

For the \textit{rising} model, we reduced the central mass concentration by extending the bulge scale radius to match the spherical halo and doubled the disc scale radius. In the \textit{declining} model, we reduce the disc scale radius by a factor of 3/4. All galactic models share the same mass for each component. We show the parameters used for the galactic models in Table \ref{tab:models}. Figure \ref{fig:rotation_curves} shows the rotation curves of the three axisymmetric models.

\begin{table*}
\centering
\caption{Parameters of the galactic models}
\label{tab:models}
{
\begin{tabular}{cccccccc}
\hline \hline
Galactic Model & $M_B$ & $b_B$ & $M_D$ & $a_D$ & $b_D$ & $M_H$ & $a_H$  \\
        & [$10^{10}M_\odot$] & [kpc] & [$10^{10}M_\odot$] & [kpc] & [kpc] & [$10^{10}M_\odot$] & [kpc] \\
(1) & (2) & (3) & (4) & (5) & (6) & (7) & (8)\\
\hline 
Rising    & 1.406 & 12.0    & 8.561 & 10.636 & 0.25 & 10.709 & 12.0 \\
Flat (Allen-Santillan)   & 1.406 & 0.387 & 8.561 & 5.318  & 0.25 & 10.709 & 12.0  \\
Declining & 1.406 & 0.387 & 8.561 & 3.988  & 0.25 & 10.709 & 24.0 \\
\hline 
\end{tabular}
}
\caption*{\Col{1} Galactic model. \Col{2} Bulge mass. \Col{3} Bulge scale length. \Col{4} Disc mass. \Col{5} Disc radial scale length. \Col{6} Disc vertical scale length. \Col{7} Halo mass. \Col{8} Halo scale length.}
\end{table*}

\subsection{Bar model}
\label{sec:bar_model}

To simulate the bar potential, we use a Ferrers ellipsoid of index $n=2$ \citep{Binney1987}. The density distribution in Cartesian coordinates is:

\begin{equation}
\rho(x,y,z) =
\begin{cases}
\frac{105}{32\pi abc} M_{Bar} (1 - m^2)^2, &  m<1\\
0,  & m>1 \\
\end{cases}
\end{equation}

where $M_{Bar}$ is the bar mass and $m^2 = (x/a)^2 + (y/b)^2 + (z/c)^2$. The parameters $a, b, c$ are the semi-axes length in the $x, y$ and $z$ directions respectively, with $a > b >= c$. The forces produced by this distribution are described in \cite{Pfenniger1984}.

We introduce the bar adiabatically as a smooth function of time by transferring mass from the bulge to the bar. We used the fifth-degree polynomial described in equation 4 of \cite{Dehnen2000}, which guarantees a smooth transition to the barred state. We set the time growth of 500 Myr, similar to \cite{Romero2015}.

Orbital and dynamical studies have shown that self-consistent bars cannot extend beyond their co-rotation radius $R_{CR}$, where the bar and the disc rotate at the same angular speed. Beyond $R_{CR}$ stellar orbits change their orientation, becoming perpendicular to the bar \citep{Athanassoula1980}. Also, the density of resonances increases near $R_{CR}$ leading to chaotic behaviour in the phase space \citep{Contopoulos1980}. Because of its physical importance, the dimensionless parameter $\mathcal{R} = R_{CR}/ a $ is used to parametrize the bar rotation rate. Bars are classified kinematically as ``slow" if \Rpar{} >1.4, and ``fast" if \Rpar{}<1.4. The theoretically impossible case of \Rpar{}<1 is referred as ``ultra-fast". 

In this work, we explore the following bar parameter space: (i) Three bar radius corresponding to $a = 0.5, 1.0, 1.5$ times the disc radial scale radius of the Allen-Santillan model. (ii) Two axial ratios between the minor and major axis $b/a = 0.3, 0.6$, corresponding to prolate and oblate bars, respectively. (iii) Two bar masses that correspond to a complete and a half mass transfer between the bulge and the bar. (iv) Three values for the dimensionless parameter \Rpar, exploring the scenario of slow, fast and ultra-fast bars. We do not explore the effects of the vertical axial ratio, which was set to $c/a=0.3$ for all our simulations. Table \ref{tab:bar_parameters} shows a summary of the bar parameter space. Our simulations explore 36 combinations for the bar model, and 3 galactic potential models, yielding a total number of 108 simulations. All our simulations use 1 million test particles.

\begin{table}
\centering
\caption{Bar parameter space}
\label{tab:bar_parameters}
{
\begin{tabular}{cccc}
\hline \hline
 Parameter & Values \\
 (1) & (2) \\
\hline 

a [kpc]        & 2.659, 5.318, 7.977   \\
b/a            & 0.3, 0.6        \\
$\mathcal{R}$  & 0.9, 1.2, 1.5   \\
$M_{Bar}$ [$10^{10}M_\odot$]     & 0.703, 1.406        \\
c/a            & 0.3 \\
\hline 
\end{tabular}
}
\caption*{\Col{1} Bar parameters. From top to bottom: Length, axial ratio, rotation rate, mass and vertical axial ratio. \Col{2} Values explored.}
\end{table}

Given the bar length $a$ and the parameter \Rpar, the pattern speed \Om{} is estimated by evaluating axisymmetric angular velocity curve at the corotation resonance $R_{CR} = \mathcal{R} \cdot a$. However, as the bar is being introduced, the mass distribution and the location of the resonances changes. This is especially important in the \textit{rising} model, where the bar formation increases the central mass concentration. In those cases, we use the unstable Lagrange point $L1$ as a proxy for the location of $R_{CR} $\citep{Binney2008}. If the relative difference between $L1$ and the axisymmetric $R_{CR}$ is greater than 10\%, we correct the value of \Om{} until both distances match. From hereafter, when we refer to \Rpar{} we are using the $L1/a$ ratio, accounting for the bar mass redistribution.

\subsection{Initial conditions and orbit integration}
\label{sec:initial_conditions}


A natural space to study spirals is the $(\theta, \ln R)$ plane, where $\theta$ is the azimuthal angle and $R$ the cylindrical radius. In this space, a logarithmic spiral can be described with the straight line equation \citep{Lin1964}:

\begin{equation}
    R(\theta)  = R_0 \times e^{\theta \tan \alpha}
\label{eq:spiral}    
\end{equation}

where, $R=R_0$ at $\theta = 0$ and $\alpha$ is the pitch angle. 


However, a disc populated with test particles uniformly distributed in the $(\theta, \ln R)$ space would have few particles in the inner disc. Instead, we choose to set the initial spatial distribution to be uniform in the \thetalog{} space, which keeps the logarithmic spacing in the radius, populates the inner disc and highlights the spiral over-densities over a uniform background. The resulting density distribution resembles that of a single exponential disc. For the vertical dimension, we choose a usual $sech^2(z/z_0)$ law distribution \citep{vanderKruit1981} with scale-height $z_0=0.25$ kpc.

We generated the initial velocities using the \cite{Hernquist1993} moments method. This procedure constrains the shape of the velocity ellipsoid, by assuming the radial and vertical dispersion are proportional and constant throughout the disc $(\sigma_R^2 \propto \sigma_z^2)$. For simplicity, we set the velocity dispersion ratio to  $\sigma_z/\sigma_R = 1$. In Section \ref{sec:velocity_ellipsoid} we explore the effect of other normalisation constants with greater radial velocity dispersion for a subset of simulations. 

Before introducing the bar, we first integrate the test particles orbits in the axisymmetric potential for 3 Gyr so they relax and reach the statistical equilibrium \citep{Romero2015}. We performed the integration with a time adaptive fifth-order Runge-Kutta integrator using the \code{Fortran} subroutines \code{odeint} and \code{rkqs} \citep{Press2007}. We followed the Jacobi energy and vertical velocity distribution for a subset of test particles to confirm the stability. 

\begin{figure*}
    \centering
    \includegraphics[width=\linewidth]{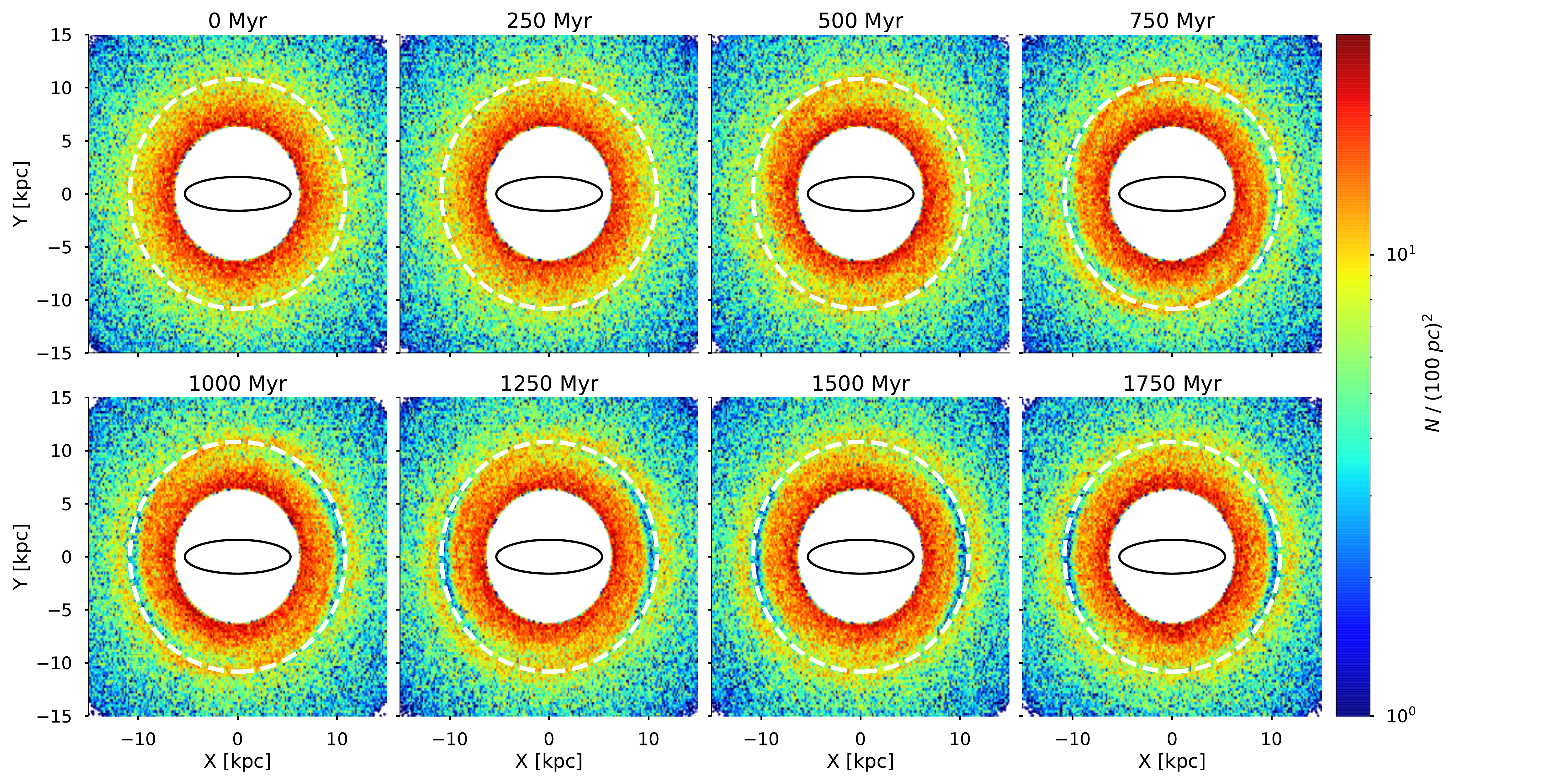}
    \caption{Snapshots of a simulation in the \textit{flat} model and bar properties $a=5.318$ kpc, $b/a = 0.3$, $\mathcal{R}=1.2$, $M_{Bar} = 1.406 \times 10^{10} M_\odot$. We masked particles inside corotation \CR{} to make spirals more visible. The OLR is shown with a white segmented line. The bar is shown with a black ellipse. The bar is introduced adiabatially by transferring mass from the bulge component up to 500 Myr. The spiral amplitude reaches a maximum at 1000 Myr, and decreases afterwards.}
    \label{fig:snapshots_xy}
\end{figure*}

\begin{figure*}
    \centering
    \includegraphics[width=\linewidth]{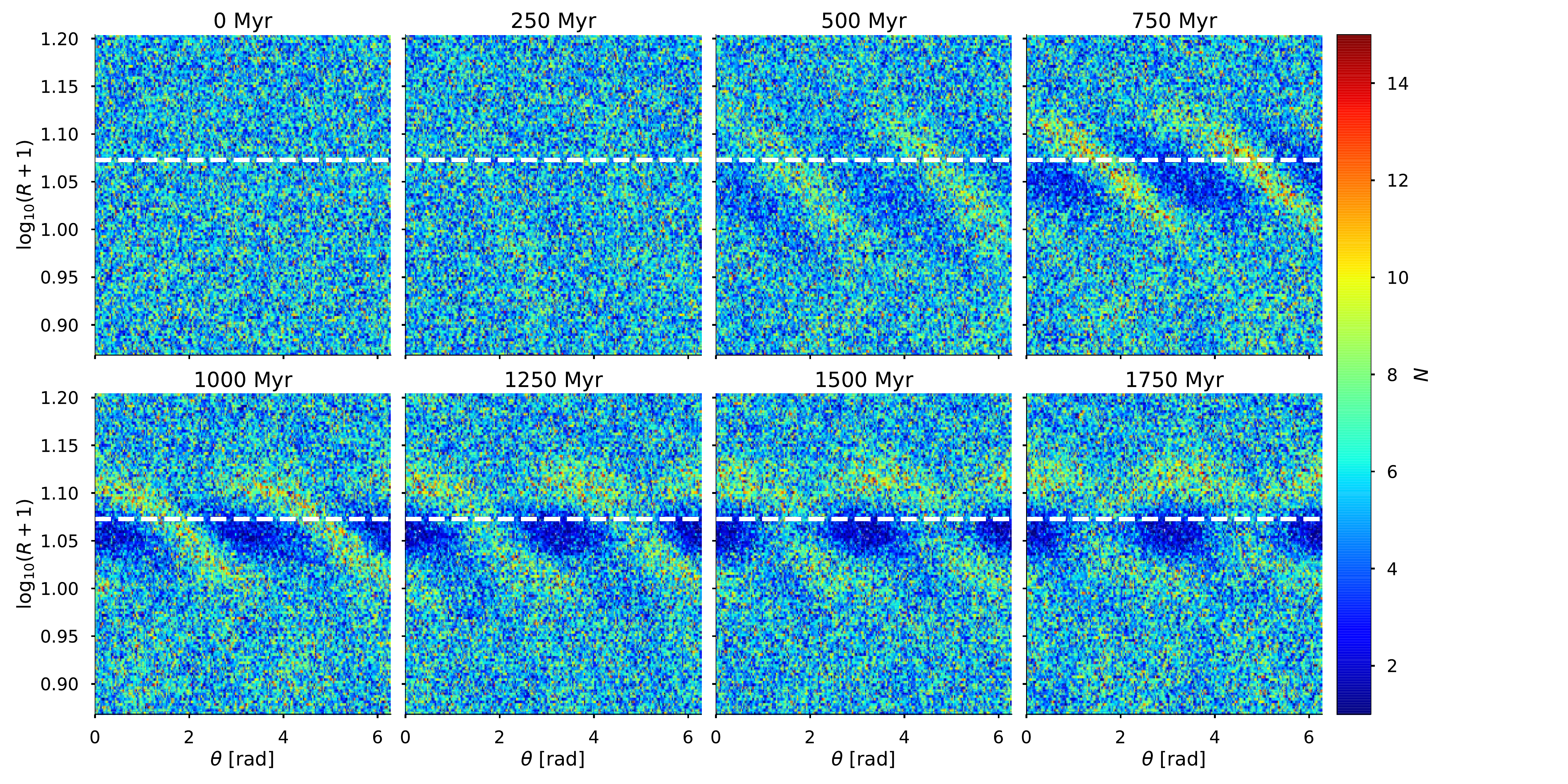}
    \caption{Same as Figure \ref{fig:snapshots_xy}, but in the \thetalog{} plane. The white segmented line highlights the location of the OLR. Notice that in the first snapshot, the particles are uniformly distributed.} 
    \label{fig:snapshots_thr}
\end{figure*}

Figures \ref{fig:snapshots_xy} and \ref{fig:snapshots_thr} show snapshots of a simulation using the \textit{flat} model and bar parameters $a=5.318$ kpc, $b/a = 0.3$, $\mathcal{R}=1.2$, $M_{Bar} = 1.406 \times 10^{10} M_\odot$ in the $(x,y)$ and \thetalog{} planes, respectively. Both figures are in the bar rotating reference frame. The snapshots are separated in time steps of 250 Myr. The first snapshot corresponds to the relaxed disc distribution. The bar stops growing in mass at 500 Myr. We show the corotation resonance with a white solid line. The Inner and Outer Lindblad Resonances (ILR and OLR from hereafter) are shown with white segmented lines. These resonances were estimated using the axisymmetric angular velocity curve and the epicyclic frequency curve.

\section{Estimating the properties of the spiral arms}
\label{sec:DBSCAN_spiral_properties}

\subsection{Detecting over-densities using DBSCAN}

To detect the spiral over-densities, we use the density-based clustering algorithm \code{DBSCAN} \citep{Ester1996}. The algorithm works by classifying each point in a given space as ``core", ``member" or ``noise", based on two parameters: (i) $\epsilon$ which specifies the distance between two points to be considered ``neighbours", and (ii) $mincnt$ which specifies the minimum number of neighbours a point must have to be classified as a ``core" of a cluster. All points in the $\epsilon$-neighbourhood of a core, that do not satisfy the $mincnt$ condition are classified as ``members" of the cluster. Finally, all points that do not inhabit a cluster are classified as ``noise". 

\code{DBSCAN} is especially useful to find arbitrarily shaped structures and does not require knowing a priori the number of clusters in the data. The use of the algorithm has been increasing in astronomy in recent years. To mention a few applications: identifying lensed features in residual images \citep{Paraficz2016}, classifying eclipsing binaries light curves \citep{Kochoska2017}, detecting low surface brightness galaxies in the Virgo cluster \citep{Prole2018} and finding open clusters in the Gaia data \citep{Castro-Ginard2018, Castro-Ginard2020}. 

\begin{figure*}
    \centering
    \begin{subfigure}{0.25 \textwidth}  
		\includegraphics[width =  \textwidth]{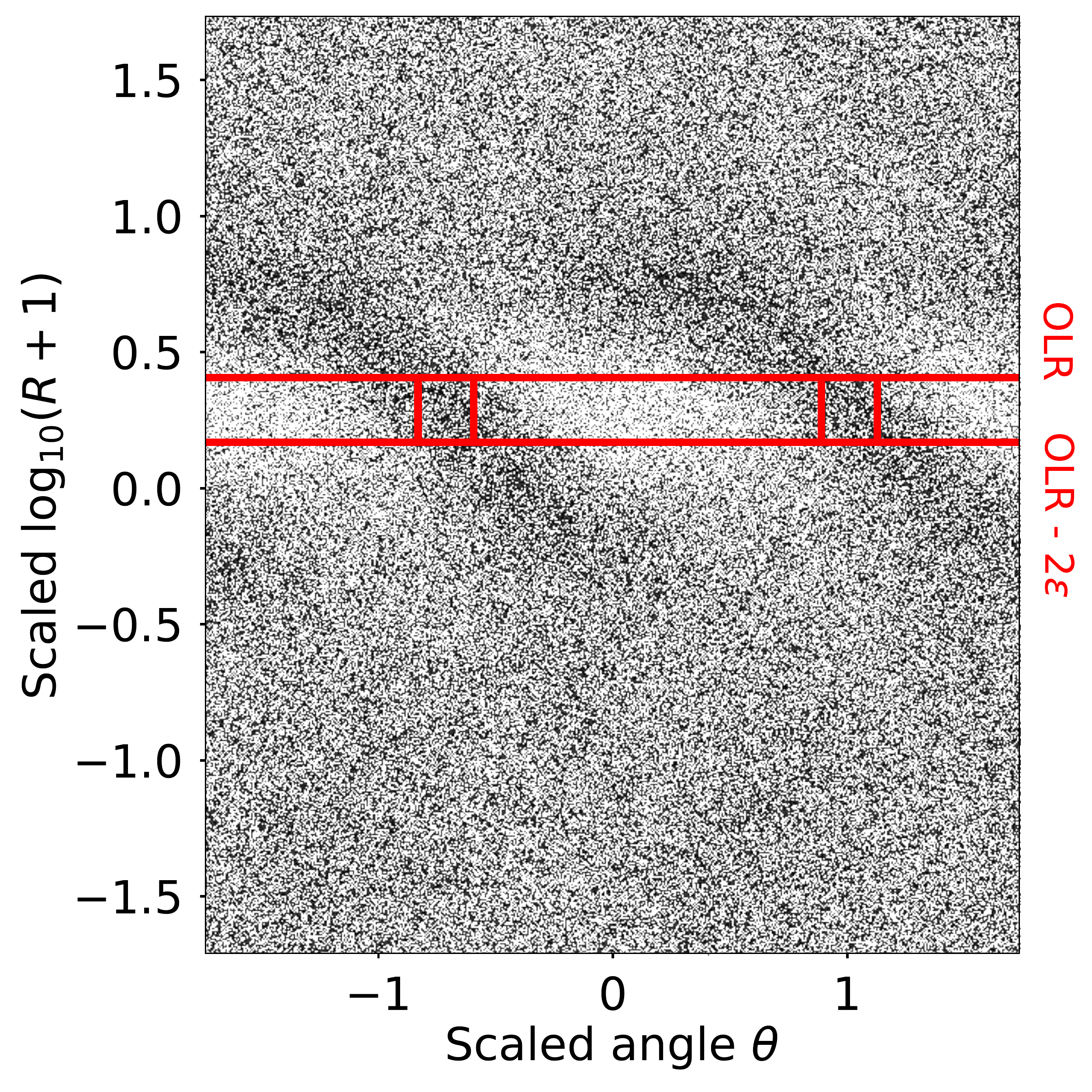}
    \end{subfigure}%
    \begin{subfigure}{0.25 \textwidth}  
		\includegraphics[width =  \textwidth]{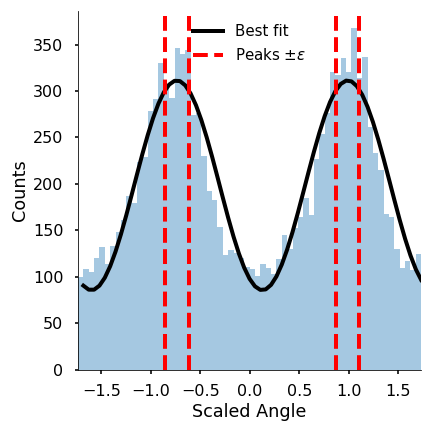}
    \end{subfigure}%
    \begin{subfigure}{0.25 \textwidth}
		\includegraphics[width =  \textwidth]{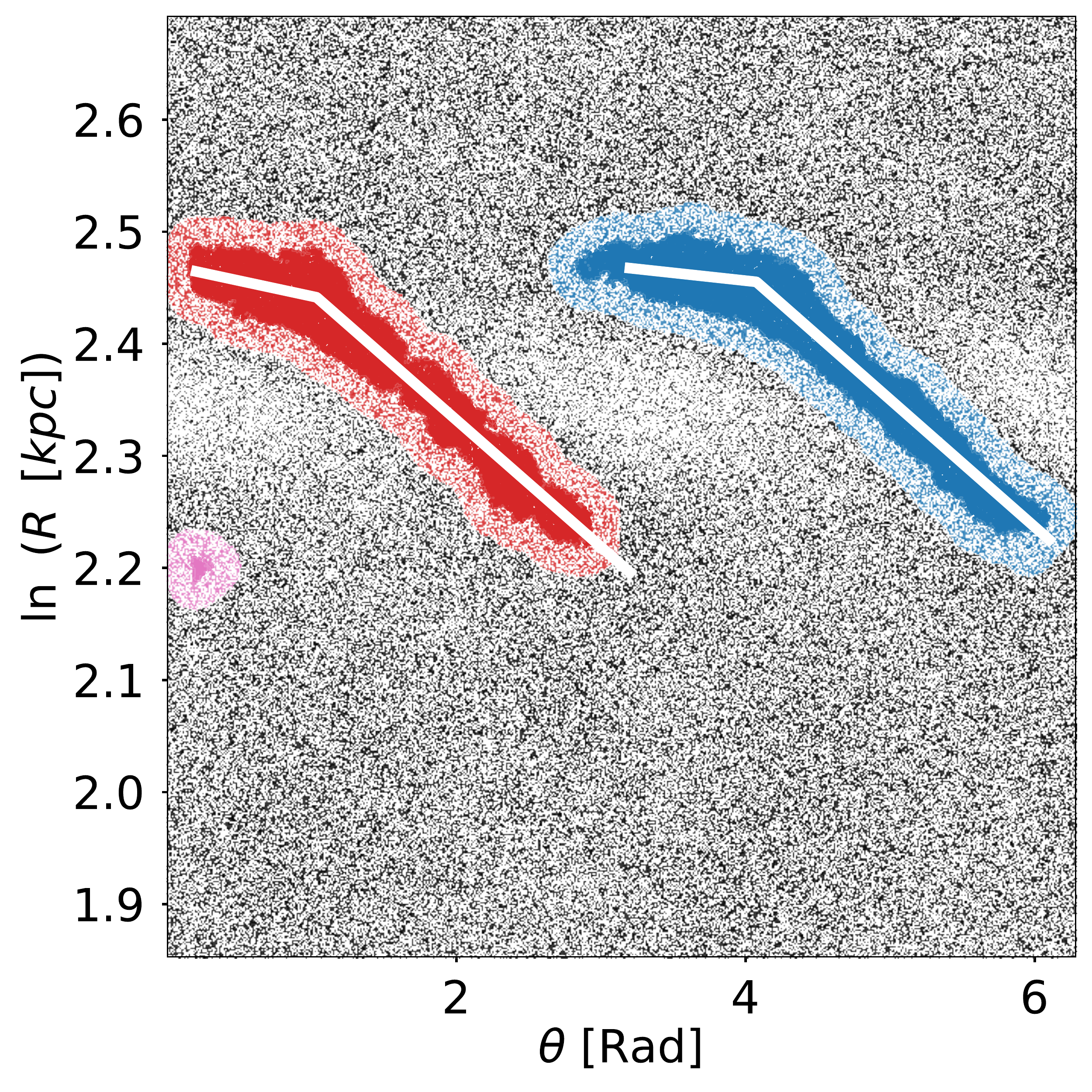}
     \end{subfigure}%
     \begin{subfigure}{0.25 \textwidth} 
		\includegraphics[width =  \textwidth]{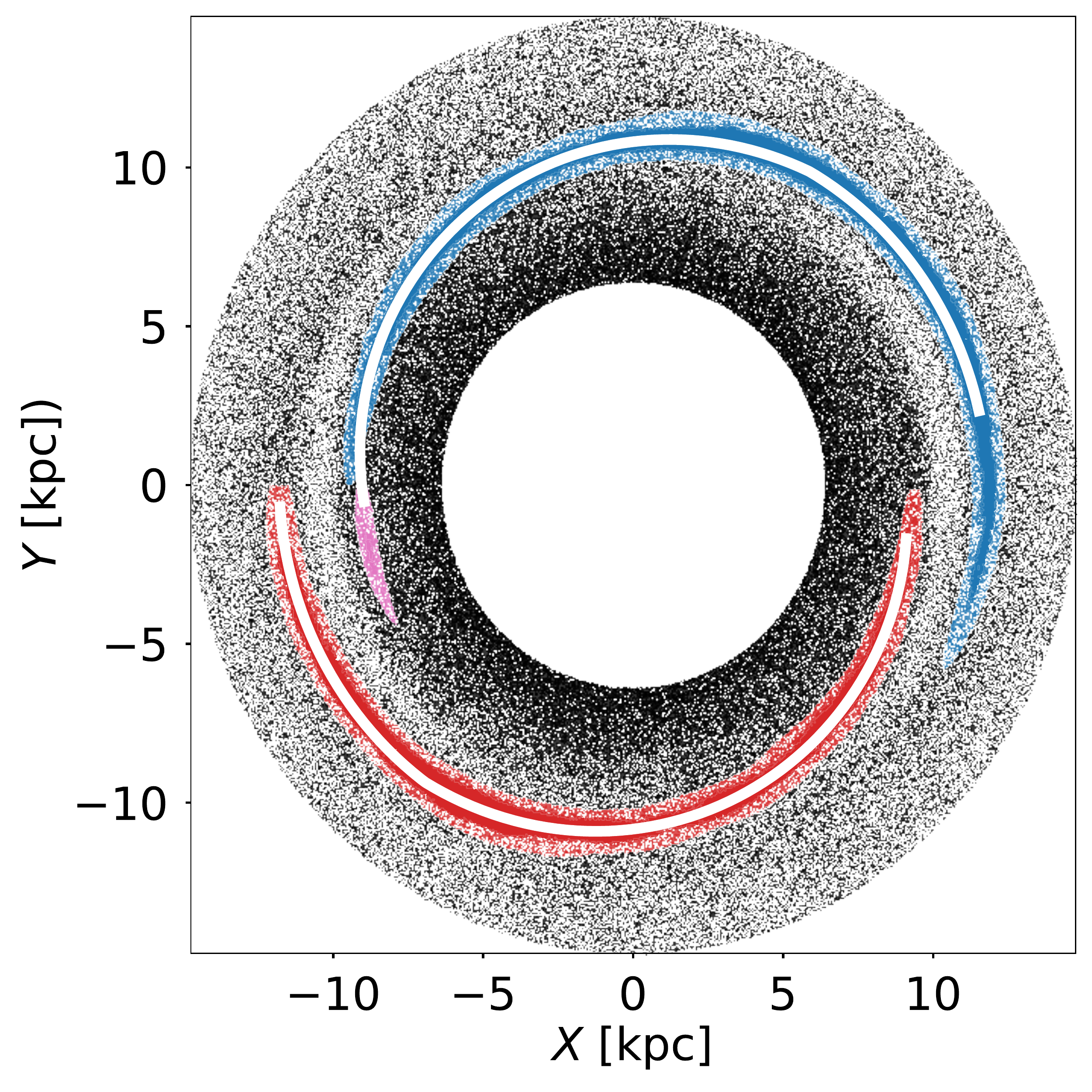}
    \end{subfigure}
    \caption{Detecting spiral over-densities with DBSCAN. We use a snapshot from the simulation with bar parameters $a=5.318$ kpc, $b/a = 0.3$, $M_{bar}=1.406 \times 10^{10} M_\odot$ , $\mathcal{R} = 1.2$, in the \textit{flat} galactic model. First panel: We select particles between \CR{} and the OLR + 4 kpc where the spiral arms are formed. We re-scale the data so distances in radius and angle are the same. We mask the particles in the range [OLR $- 2\epsilon$, OLR] shown with two red horizontal lines. At the peak spiral density we draw the two squares of side $2\epsilon$. Second panel: Distribution of the masked particles vs. the scaled angle. We fit a cosine function to get the amplitude and peaks of the spiral over-density. We count the number of test particles around the peaks $\pm \epsilon$, shown with red vertical lines. We determine $mincnt$ by multiplying this count by $\pi/8$. Third panel: Using the estimated parameters $(\epsilon, mincnt)$, \code{DBSCAN} identifies three clusters in the data. The points classified as ``core'' are shown with slightly bigger dots, so they appear as a solid coloured area. ``Member'' points surround the core area in an $\epsilon$ size envelope. Notice this plot is in the $(\theta, \ln R)$ space. We use a piecewise linear function to fit the identified clusters shown with two white lines. We measure the pitch angle using the greatest slope in the piecewise fit. Fourth panel: Same as the third panel, but in the (x,y) plane. }
    \label{fig:DBSCAN_density}
\end{figure*}

Choosing meaningful parameters $(\epsilon, mincnt)$ is challenging when the size and density of the clusters are unknown. The parameter $\epsilon$ is related to the spatial resolution of the clusters and can be determined from their expected size. The spiral arms produced in our simulations typically have widths of $\sim 0.6$ kpc, so we choose to fix the value of $\epsilon$ to 0.3 kpc. The parameter $mincnt$ is related to the expected density, or in this case, the spiral amplitude. However, the amplitude changes with time and between simulations. We performed the following steps to determine an appropriate value of $mincnt$ at any given snapshot. We show an example of the procedure in Figure \ref{fig:DBSCAN_density}:

\begin{enumerate}[i]
    \item We mask all particles between \CR{} and OLR + 4 kpc, where we expect the spiral arms to be located. We re-scale the data, so the distances in $\log (R +1)$ and $\theta$ are the same in an Euclidean metric. We also re-scale the value of $\epsilon$. The first panel of Figure \ref{fig:DBSCAN_density} shows a snapshot of a simulation where the spiral arms have already formed, and the data has been re-scaled.
    \item A second mask is used to select particles that lie in the radial range $R =  [OLR - 2 \epsilon, OLR]$. In the first panel of Figure  \ref{fig:DBSCAN_density}, the two red horizontal lines delimit this range.
    \item All our simulations produced two symmetrical spirals arms. We bin the masked particles in 64 bins and fit a simple cosine wave function $f(\theta) = A \cos(2 \theta + \phi) + C$. This is equivalent to the $m=2$ Fourier component of the azimuthal light profile near the OLR. We use the normalised amplitude of the fit, $A/C$, as a proxy of the spiral amplitude. In the second panel of Figure \ref{fig:DBSCAN_density}, we show the masked particles distribution and the cosine fit.
    \item We count the number of particles around the peaks $\pm \epsilon$, illustrated with two red segmented lines in the second panel of Figure \ref{fig:DBSCAN_density}. The resulting count is the expected number of particles in two squares of size $2 \epsilon$ at the peak of the spirals. We show these squares in the first panel of Figure \ref{fig:DBSCAN_density}. 
    \item We multiply the resulting count by $\pi/8$ which is the ratio between the area of a circle with radius $\epsilon$ and two squares of side $2\epsilon$. This results in our estimation for $mincnt$. 
\end{enumerate}

Once we have estimated $(\epsilon, mincnt)$, \code{DBSCAN} can detect the spiral over-densities as shown in the third and fourth panels of Figure \ref{fig:DBSCAN_density}. In this example, the algorithm detected 3 clusters, shown in different colours. We plot the core points with slightly bigger dots, so they appear as a solid coloured region. Member points surround the core with an envelope of radius $\epsilon$. Notice that the pink cluster is just a continuation of the blue spiral, but displaced by $2 \pi$ radians. In such cases, we join the separated clusters by manually adding the $2 \pi$ rotation. 

Not every simulation produces spirals as clearly as in the example shown in Figure \ref{fig:DBSCAN_density}. Some models produce very weak spiral arms that are almost indistinguishable from the background. In those cases, we reduce the value of $mincnt$ manually until the algorithm can detect the underlying spiral structure.

\subsection{Estimating the pitch angle}

Although we identify the spiral over-densities in the $(\theta, \log_{10} R+1)$ plane, the measurement of the pitch angle is done in the $(\theta, \ln R)$ space, as described by equation \ref{eq:spiral} (i.e $\arctan$ of the slope of the over-densities). In some cases, the spirals wind up at the outer radius, forming a ring-like structure just outside of the OLR. These kind of rings are common in numerous simulations with test particles \citep{Schwarz1981, Schwarz1984, Bagley2009} and barred galaxies \citep{Buta1991OLR, Buta1996, Buta2017_Rings_GalaxyZoo}. The snapshot shown in Figure \ref{fig:DBSCAN_density} is an example of such behaviour. To separate the rings from the spiral, we fit a piecewise linear function and use the slope of the spiral segment to estimate the pitch angle. The fitted piecewise function is shown with white lines in the third and fourth panels of Figure \ref{fig:DBSCAN_density}. Notice how the change in pitch angle cannot be distinguished `by eye' in the $(x,y)$ plane. We measure the pitch angle in both spirals using the whole cluster and only core points. From hereafter, when we refer to the pitch angle we are using the average from these measurements.

\subsection{Estimating the spiral amplitude}

\begin{figure}
    \centering
    \includegraphics[width=\linewidth]{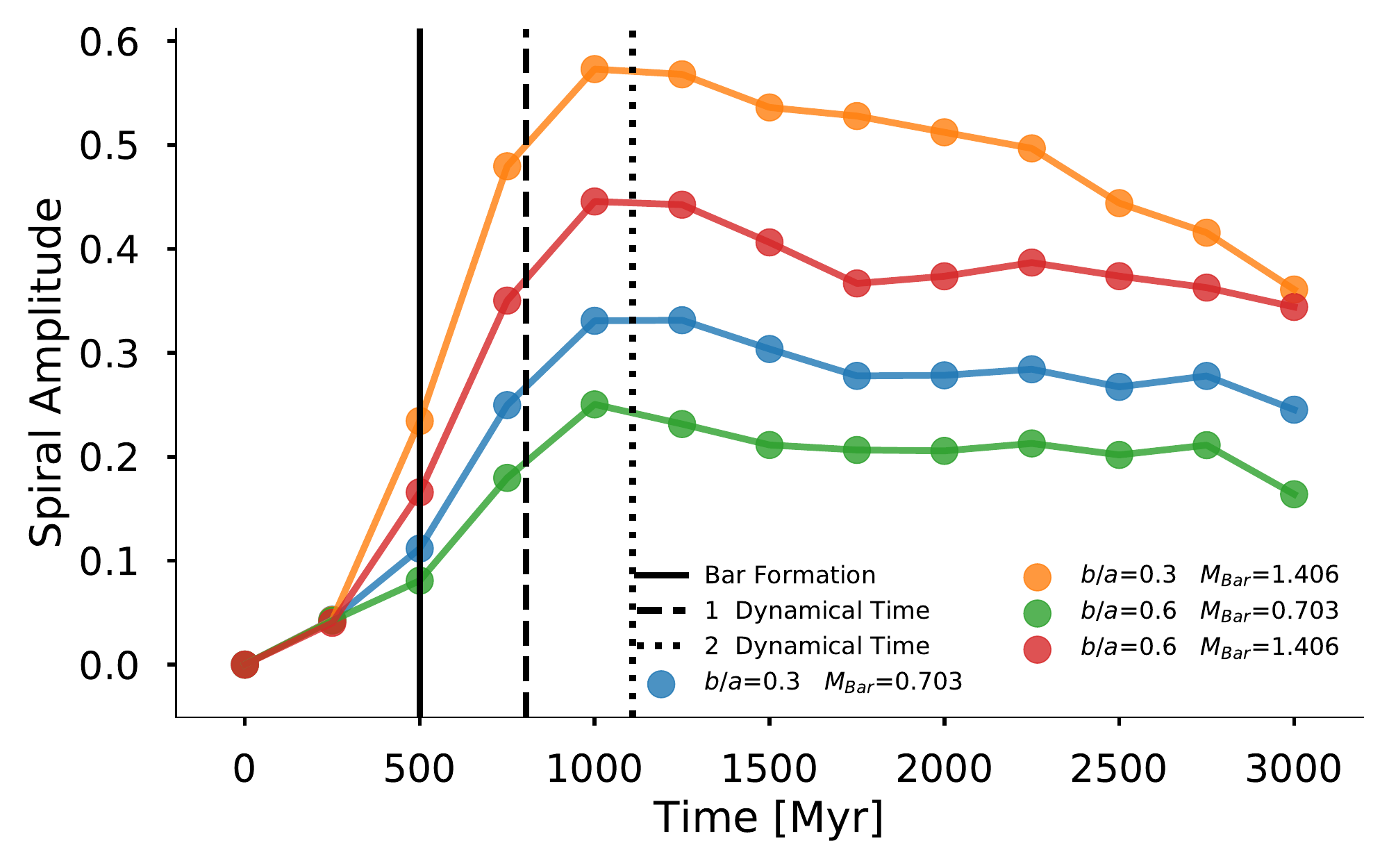}
    \caption{Spiral arm amplitude as a function of time, for the family of simulations with bar parameters $a = 5.318$ kpc and $\mathcal{R} = 1.2$ in the \textit{flat} model. The solid black line shows the bar formation (500 Myr), the segmented and dotted lines show 1 and 2 dynamical times at the OLR after the bar formation, respectively.}
    \label{fig:Amplitude}
\end{figure}

We use the normalised amplitude of the fitted cosine wave function (second panel in Figure \ref{fig:DBSCAN_density}) as a proxy of the spiral amplitude. In general, all our simulations showed the same trend: As the bar is being introduced, the spiral amplitude increases as a function of time until it reaches a maximum and slowly decreases. In Figure \ref{fig:Amplitude} we show the spiral amplitude as a function of time of four simulations in the \textit{flat} galaxy model that share the same bar length $a$ and parameter \Rpar{} (and thus, the same \Om{}), but vary in mass and axis ratio. In most simulations, the amplitude peaks between 1 or 2 dynamical times at the OLR after the bar has formed. After 3 to 4 dynamical times the spiral structure becomes more diffuse. Simulations that share $a$ and \Rpar{}, usually have spirals that peak in the same snapshot, except for some simulations in the \textit{rising} model, that are more sensitive to the bar potential. 

In the next section we discuss which bar parameters produce stronger spirals, but from this figure is clear that the mass and shape of the bar are strongly co-related to the spiral amplitude. 

\section{Relations between the bar parameters and the spiral properties}
\label{sec:bar-spiral}

The spiral arms produced in our simulations are the collective result of the bar perturbing the stellar epicyclic orbits. Thus, our spiral arms are produced by the density-wave mechanism and do not account for the effects of self-gravity. Our results should be interpreted as the initial spiral perturbation that arises from the presence of different kinds of bars. Spiral arms produced this way, are expected to form near the OLR \citep{Athanassoula1980}, and be tightly wound with small pitch angles.

Because of the great variety of parameters we are exploring, the spirals arms form at different times, amplitudes and locations throughout the disc. To make the comparison between different simulations as fair as possible, we use the measurements of the snapshot where spirals are at their greatest amplitude.

\subsection{Observed correlations}

\begin{figure*}
    \centering
    \includegraphics[width=\linewidth]{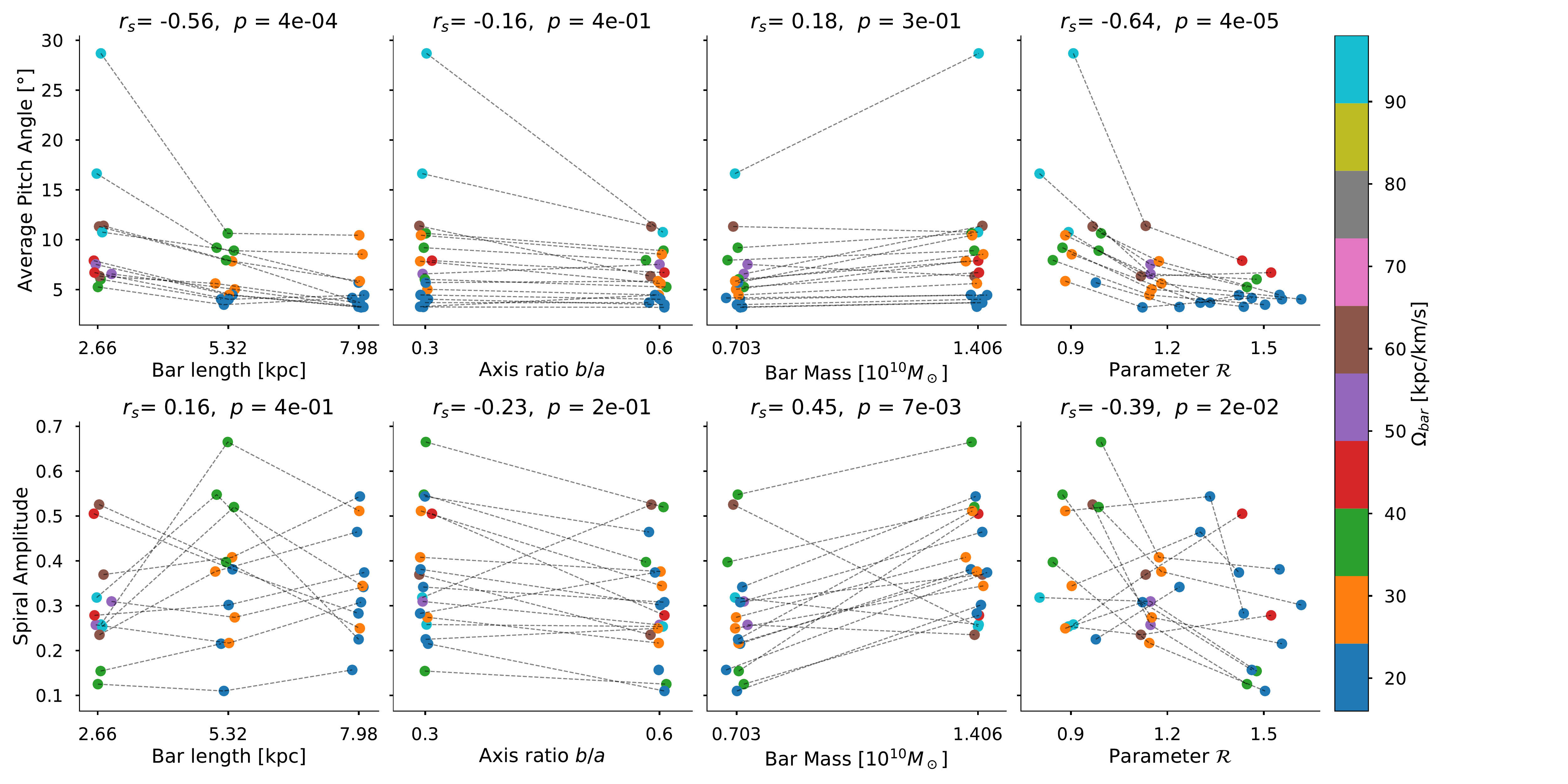}
    \caption{Spiral properties vs bar parameters in the \textit{rising} galactic model. Top row: Relations with the average pitch angle. Bottom row: Relations with the spiral amplitude. The grey segmented lines join simulations that share the same bar parameters, except for the one that is being plotted. The Spearman correlation coefficient and the corresponding p-value are shown at the top of each panel.}
    \label{fig:Correlations_AS}
\end{figure*}

\begin{figure*}
    \centering
    \includegraphics[width=\linewidth]{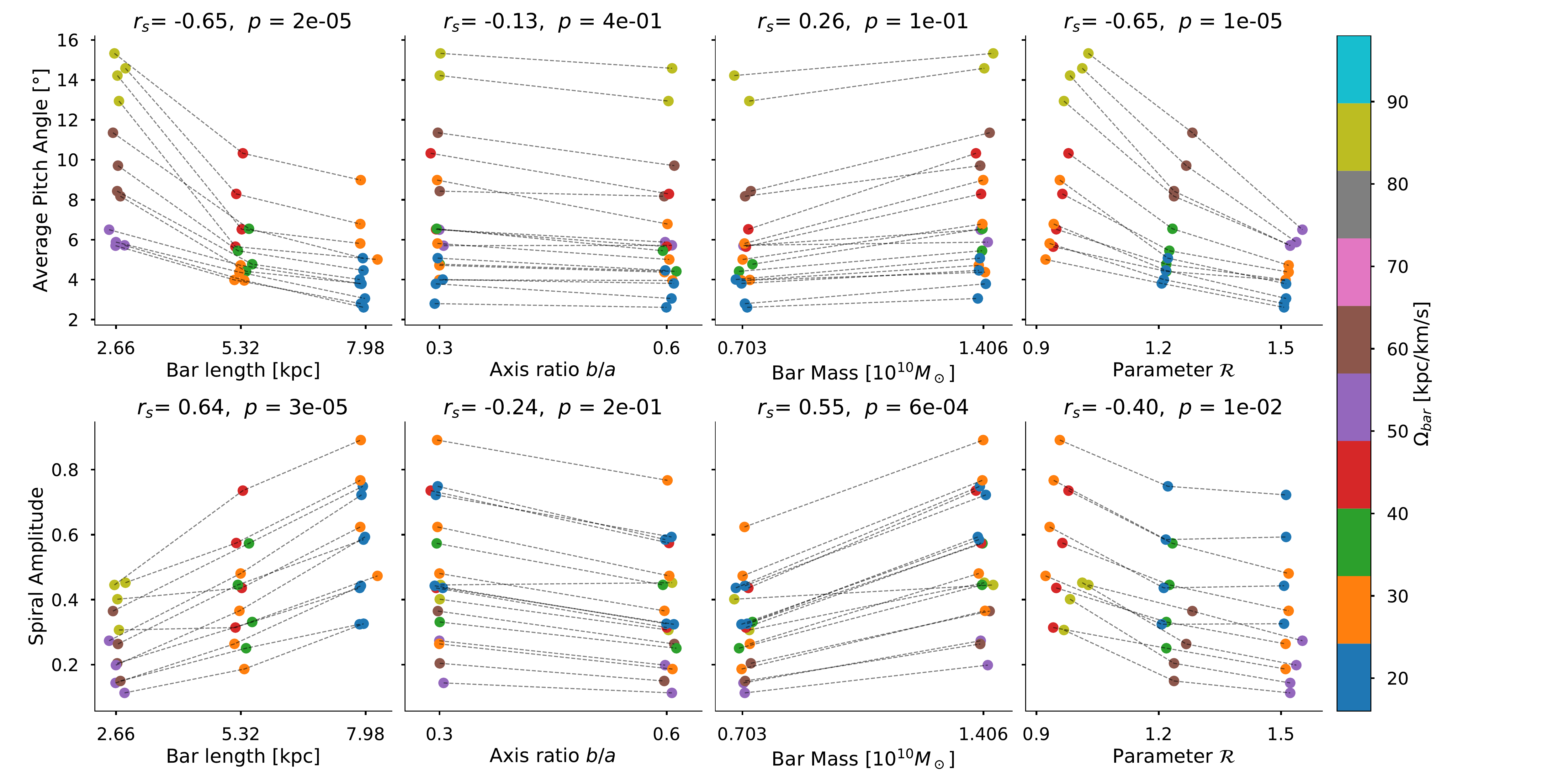}
    \caption{Same as Figure \ref{fig:Correlations_AS}, but in the \textit{flat} model.}
    \label{fig:Correlations_RC}
\end{figure*}

\begin{figure*}
    \centering
    \includegraphics[width=\linewidth]{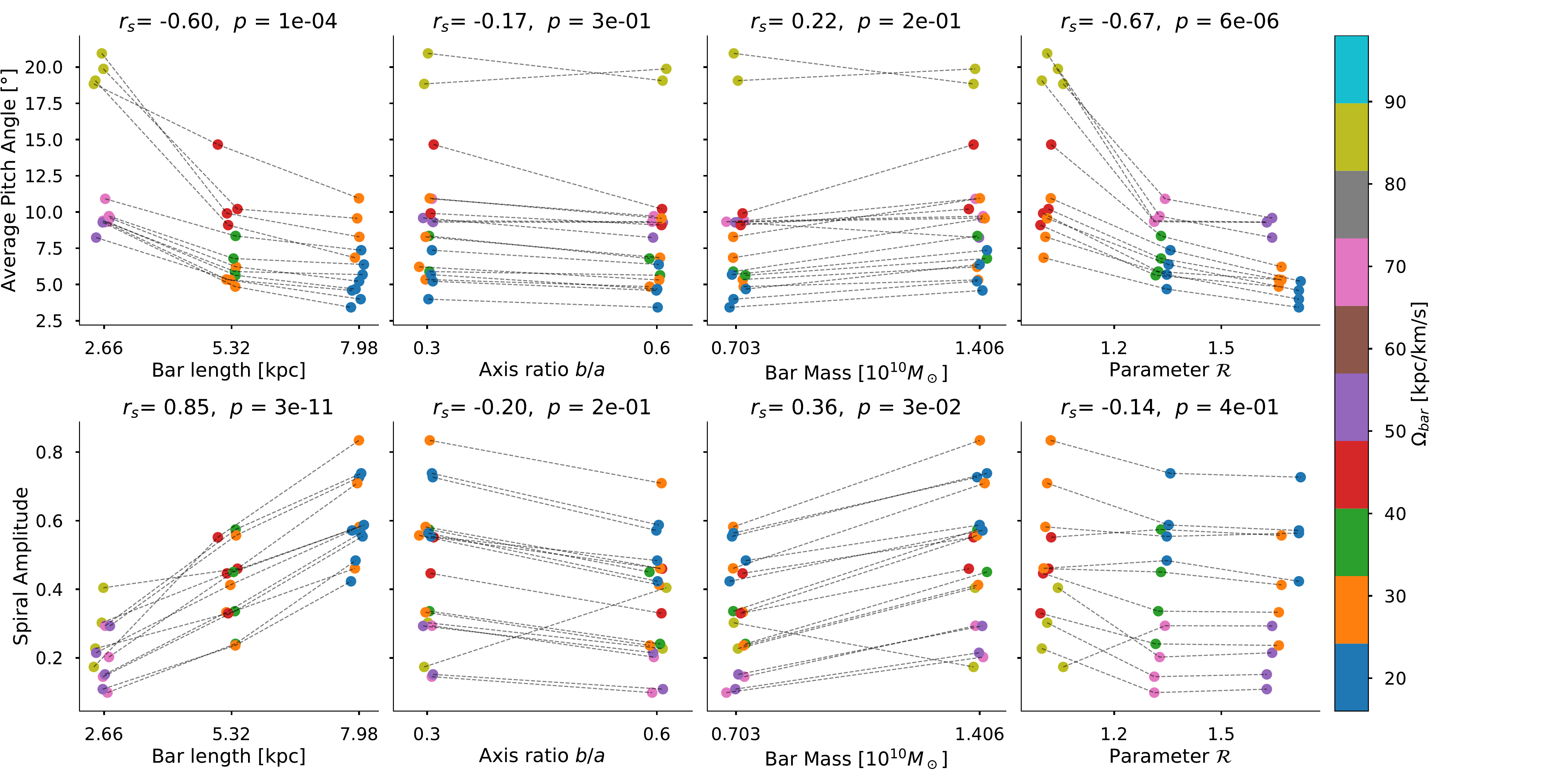}
    \caption{Same as Figure \ref{fig:Correlations_RC} and \ref{fig:Correlations_AS}, but in the \textit{declining} model. }
    \label{fig:Correlations_DC}
\end{figure*}

In Figures \ref{fig:Correlations_AS}, \ref{fig:Correlations_RC}, and \ref{fig:Correlations_DC} we show the average pitch angle and the spiral amplitude vs. the bar parameters of our simulations in the \textit{rising}, \textit{flat}, and \textit{declining} galactic models, respectively. We join with grey dotted lines the simulations that share the same bar parameters except for the one that is being plotted. The Spearman correlation coefficient $r_S$ and the corresponding statistical significance $p$ are shown at the top of each panel. We colour the results with the bar pattern speed. To facilitate the interpretation of our results, Table \ref{Tab:correlations} summarises all Spearman correlation coefficients that we mention through the text.

In all galactic models we found a strong anti-correlation of the pitch angle with the bar length $(r_S \sim -0.6)$ and the $\mathcal{R}$ parameter $(r_S \sim -0.65)$. At a first approximation, it would appear that the mass and shape of the bar do not correlate with the pitch angle, with a small $r_S$ coefficient and p-values $>0.05$. However, the connected models display a clear downward trend with $b/a$ and an upper trend with $M_{Bar}$. We performed a Student's t-test to see if these slight differences between connected simulations could happen by random chance. We could reject the null hypothesis in both cases in all galactic models ($p \sim 1\times 10^{-4}$ for $b/a$ and  $p \sim 1\times 10^{-5}$ for $M_{Bar}$).

The relations with the spiral amplitude does seem to depend more on the galactic models. The correlation with the bar length is non-existent in the \textit{rising} model ($r_S =0.16$), but becomes strong in the \textit{flat} and \textit{declining} models ($r_S=0.64$ and $r_S=0.85$, respectively). In contrast, the relation with the \Rpar{} parameter does not seem significant in the  \textit{declining} model ($r_S = -0.14$)  but becomes a weak anti-correlation in the \textit{rising} and \textit{flat} models ($r_S\sim-0.40$ in both cases). The connected simulations in all three models show a clear upward trend with $M_{Bar}$ and a downward trend with the ratio $b/a$, which we were able to confirm with the Student t-test, rejecting the null-hypothesis. The strong trends with the bar mass and axis ratio can also be observed in amplitude vs. time plot in Figure \ref{fig:Amplitude}.

Some of the observed correlations are related in the form of a third relationship. For example, the bar length $a$ and parameter \Rpar{} are intimately related with \Om{} and the disc rotation curve V(R):

\begin{equation}
\Omega_{Bar} = V(R_{CR}) / R_{CR} = V(R_{CR}) / ( \mathcal{R} \cdot a)
\end{equation}

\begin{figure*}
    \centering
    \includegraphics[width=\linewidth]{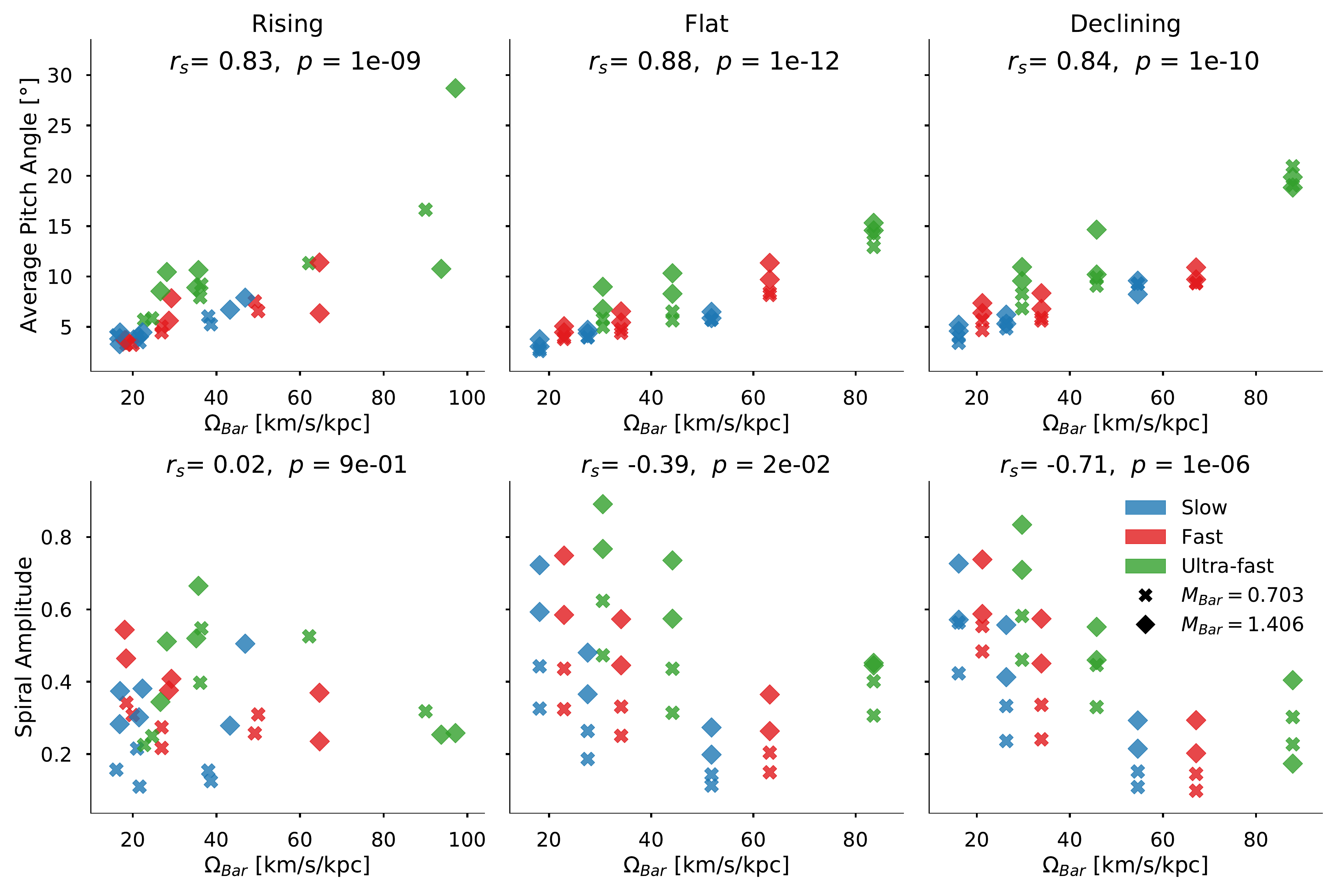}
    \caption{Pitch angle and spiral amplitude vs. bar pattern speed in the three galactic models (\textit{Rising} in the left column, \textit{Flat} in the middle, and \textit{Declining} in the right). Simulations are colour-coded according to their \Rpar{} parameter (blue for slow, red for fast and green for ultra-fast bars). The more massive bars are shown with diamonds, while their less massive counterpart are shown with crosses.}
    \label{fig:Omega_correlations.pdf}
\end{figure*}

Thus, the strong trends we observe among the pitch angle, the bar length and \Rpar{} are probably a consequence of the much stronger correlation between the pitch angle and the bar pattern speed, $(r_S \sim  0.85)$ in all three galaxy models. The relation between the spiral properties and \Om{} is shown in Figure \ref{fig:Omega_correlations.pdf} for the three galactic models. The colours and shape of the dots are used to distinguish models in \Rpar{} and $M_{Bar}$ respectively. The relation between the spiral amplitude and $\Omega_{Bar}$ depends on the galactic model, as this relation is non-existent in the \textit{rising} model, but becomes a strong anti-correlation in the \textit{declining} model. 0'p0'po

The bar length $a$, mass $M_{Bar}$ and axial ratio $a/b$ are related to the bar quadrupole moment (hereafter \QBar). For a Ferrers ellipsoid of index $n$ \QBar is:

\begin{equation}
   Q_{Bar} = M_{Bar} a^2 \left(1 - \frac{b^2}{a^2} \right) / \left( 5 + 2n\right)
\end{equation}

\begin{figure*}
    \centering
    \includegraphics[width=\linewidth]{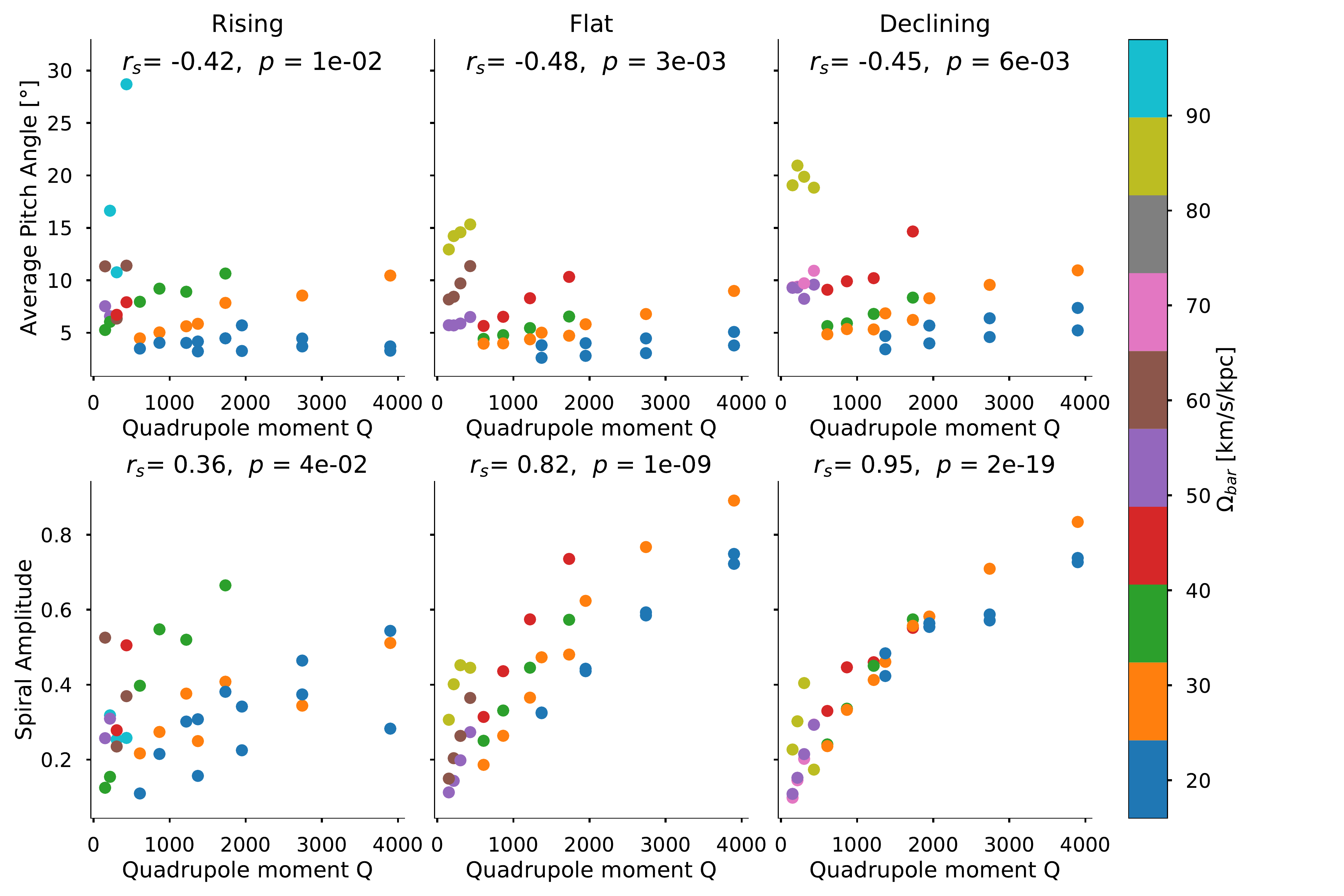}
    \caption{Pitch angle and spiral amplitude vs. bar quadrupole moment in the three galactic models.}
    \label{fig:Quadrupole_correlations}
\end{figure*}

Same as with other bar properties, in Figure \ref{fig:Quadrupole_correlations} we show the correlations between the spiral properties and the quadrupole moment in the three galactic models. Our results show a remarkably strong correlation between the spiral amplitude and \QBar{} in the \textit{flat} and \textit{declining} models. Thus, the observed correlations between spiral amplitude and  $a$, $b/a$ and $M_{Bar}$ could be a consequence of the more stronger correlation with \QBar{}. Nonetheless, this correlation does not seem to be as important in the \textit{rising} model (as with $a$, $b/a$ and $M_{Bar}$). 

The relationship between \QBar{} and the pitch angle is more complex. The general trend is negative $(r_S \sim -0.45$ in the three models). However, if we look only at simulations with the same pattern speed as shown by the colour code (i.e. simulations with same bar length and \Rpar{} parameter) the positive relation with the mass (and anti-correlation with $b/a$) becomes visible. 

\begin{table*}
\caption{Spearman correlation coefficient and statistical significance with the spiral properties}
\label{Tab:correlations}
\begin{tabular}{cccc|ccc}
\hline \hline
 & & Pitch Angle & & & Amplitude & \\
 Parameter & Rising & Flat & Declining & Rising & Flat & Declining  \\
 (1) & (2) & (3) & (4) &(5) & (6) & (7) \\
\hline
$a$       & $-0.56$ \Pval{4}{-4} & $-0.65$ \Pval{2}{-5} & $-0.60$ \Pval{1}{-4} & $0.16$ \Pval{4}{-1}  & $0.64$ \Pval{3}{-5} & $0.85$  \Pval{3}{-11}  \\
$b/a$     & $-0.16$ \Pval{4}{-1} & $-0.13$ \Pval{4}{-1} & $-0.17$ \Pval{3}{-1} & $-0.23$ \Pval{2}{-1} & $-0.24$ \Pval{2}{-1} & $-0.20$  \Pval{2}{-1} \\
$M_{Bar}$ & $0.18$ \Pval{3}{-1}  & $0.26$ \Pval{1}{-1}  & $0.22$ \Pval{2}{-1}  & $0.45$ \Pval{7}{-3}  & $0.55$ \Pval{6}{-4} & $0.36$ \Pval{3}{-2} \\
\Rpar{}   & $-0.64$ \Pval{4}{-5} & $-0.65$ \Pval{1}{-5} & $-0.67$ \Pval{6}{-6} & $-0.39$ \Pval{2}{-2} & $-0.40$ \Pval{1}{-2} & $-0.14$ \Pval{4}{-1} \\
\Om{}     & $0.83$ \Pval{1}{-9}  & $0.88$ \Pval{1}{-12}  & $0.84$ \Pval{1}{-10}  & $0.02$ \Pval{9}{-1}  & $0.39$ \Pval{2}{-2} & $0.71$ \Pval{1}{-6}  \\
\QBar{}   & $-0.42$ \Pval{1}{-2} & $-0.48$ \Pval{3}{-3} & $-0.45$ \Pval{6}{-3} & $0.36$ \Pval{4}{-2}  & $0.82$ \Pval{1}{-9} & $0.95$ \Pval{2}{-19}  \\
Shear     & $-0.44$ \Pval{1}{-2} & $-0.82$ \Pval{8}{-10} & $-0.75$ \Pval{1}{-7} & $0.00$ \Pval{1}{0} & $0.44$ \Pval{7}{-3} & $0.73$ \Pval{4}{-7}  \\
\hline
\end{tabular}
\caption*{\Col{1} Parameter. \Col{2} Pitch angle in the \textit{rising} model. \Col{3} Pitch angle in the \textit{flat} model. \Col{4} Pitch angle in the \textit{declining} model. \Col{5} Spiral amplitude in the \textit{rising} model \Col{6} Spiral amplitude in the \textit{flat} model \Col{7} Spiral amplitude in the \textit{declining} model.}
\end{table*}

\section{Influence of the rotation curve}
\label{sec:rotation_influence}

\begin{figure}
    \centering
    \includegraphics[width=\linewidth]{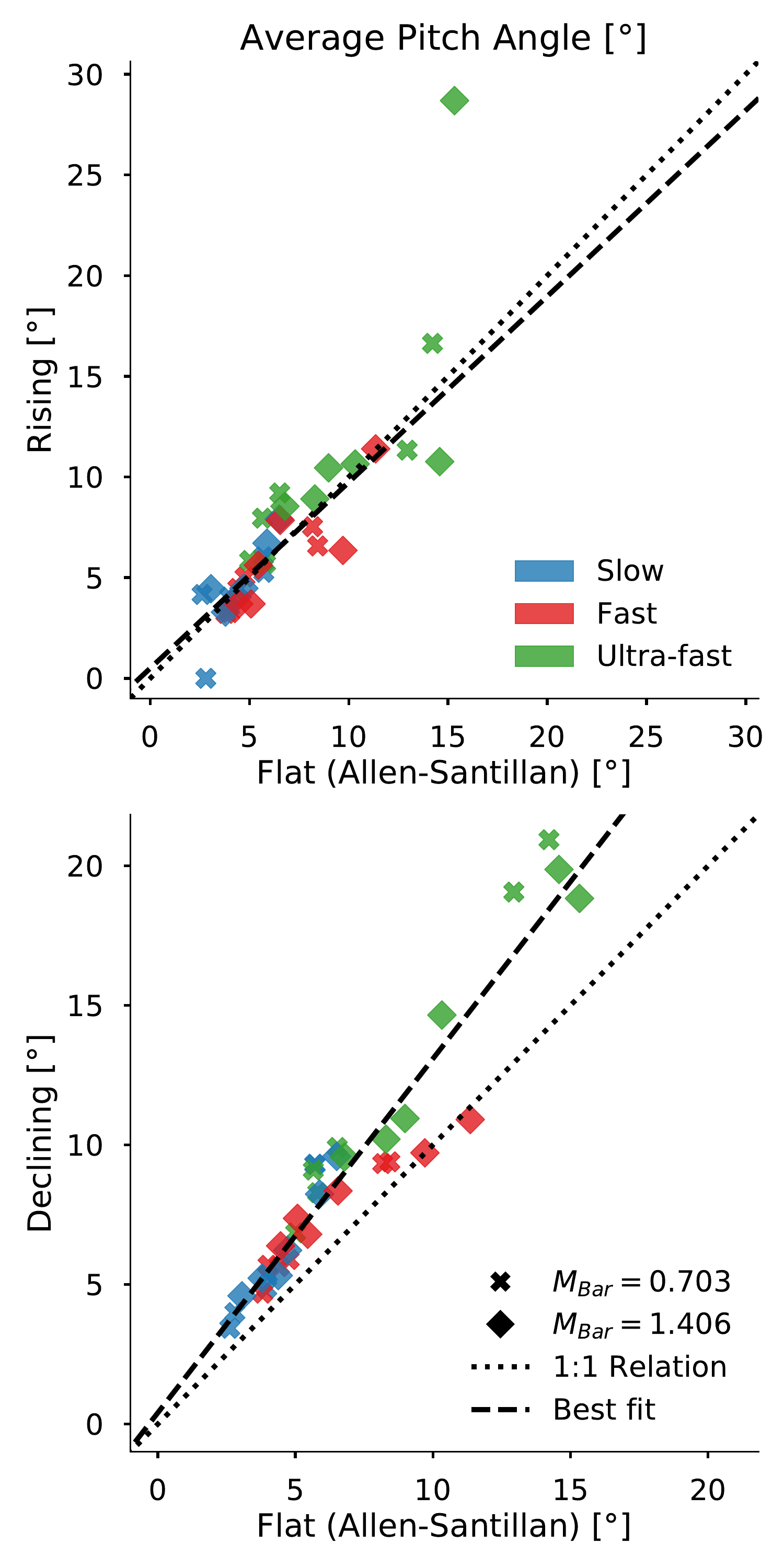}
    \caption{Comparison of the pitch angle between the three galactic models. For reference, the slopes of the best fits are $m= 0.92$  in the top panel (excluding the outlier) and $m=1.26$ in the bottom one.}
    \label{fig:Comparing_pitch}
\end{figure}

We study the effects of the rotation curve in the spiral properties, by comparing simulations with the same bar parameters, but different galaxy model. In Figure \ref{fig:Comparing_pitch} we show the pitch angles of the \textit{flat} models vs. the \textit{rising} and \textit{declining} models. As in Figure \ref{fig:Omega_correlations.pdf} the colours and shape are used to distinguish the \Rpar{} and $M_{Bar}$ respectively. We do not observe a significant difference between the \textit{rising} and \textit{flat} models, except for one outlier that corresponds to an ultra-fast, small, massive bar. In comparison, the \textit{declining} simulations produce spirals with consistently larger pitch angles. The effect also appears to be more significant with the ultra-fast bars, as those differ more significantly from the 1:1 relation. We include the best linear fit for reference. 

\begin{figure}
    \includegraphics[width=\linewidth]{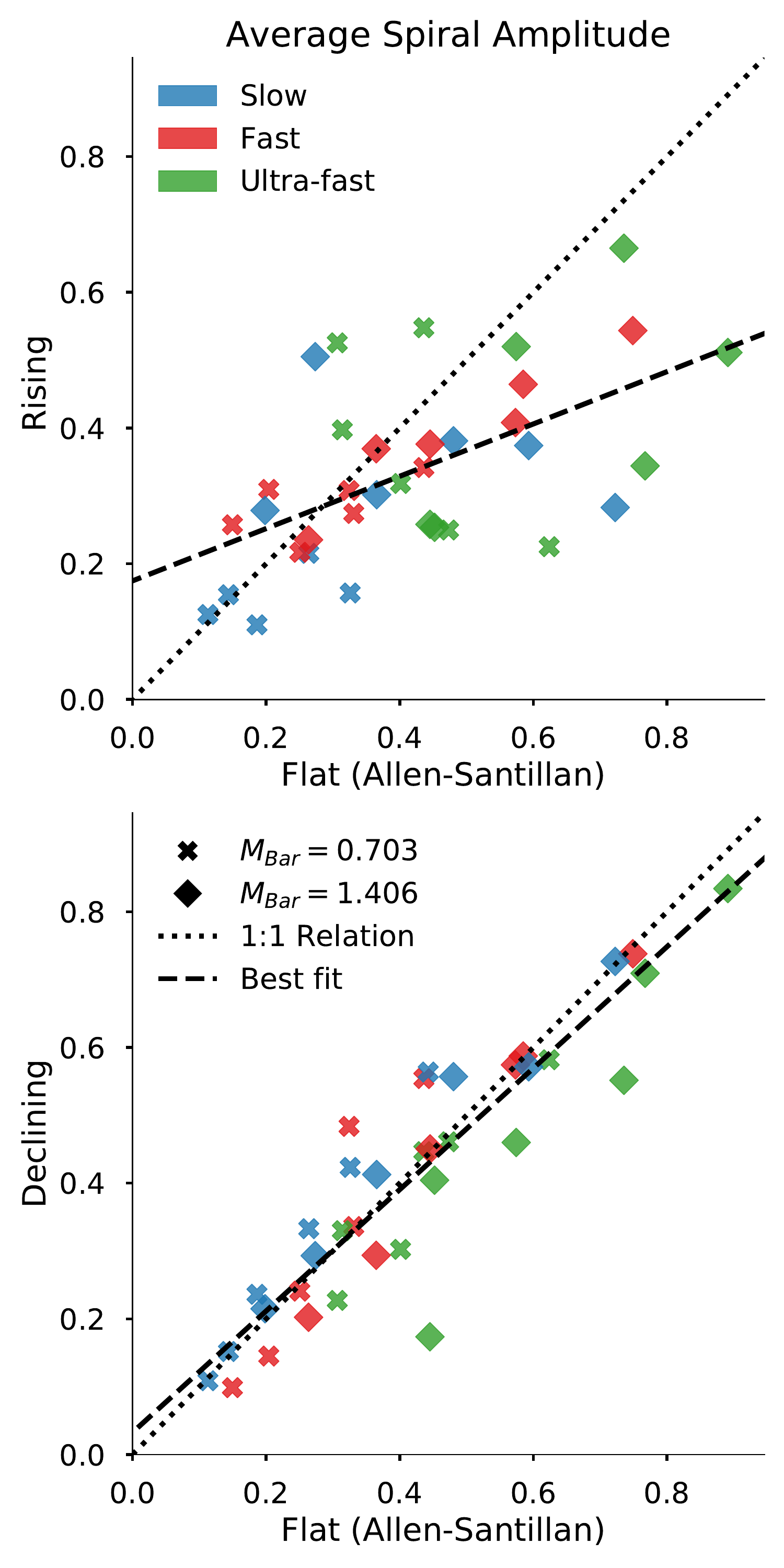}
    \caption{Comparison of the spiral amplitude between the three galactic models. For reference, the slopes of the best fits are $m= 0.39$ in the top panel and $m=0.90$ in the bottom one.}
    \label{fig:Comparing_amplitudes}
\end{figure}

Similarly, in Figure \ref{fig:Comparing_amplitudes} we compare the spiral amplitude between the three galactic models. The \textit{rising} model tends to produce weaker spirals. The difference becomes more significant with the more massive bars. On the other hand, the \textit{declining} and \textit{flat} models have similar spiral amplitudes for all the bar models.

We also estimate the shear rate $S$ (equation \ref{eq:Shear}) at the OLR in all our simulations. Values of $S<0.5$ correspond to a rising region in the rotation curve, $S=0.5$ to a flat region and $S>0.5$ to a declining region. The presence of the bar changes the shape of the rotation curve from the axisymmetric model. We estimate the perturbed rotation curve by averaging $V(R) = \sqrt{R d\phi/dR}$ over 10 angles uniformly distributed between 0 and $\pi/2$ radians. 

\begin{figure*}
    \includegraphics[width=\linewidth]{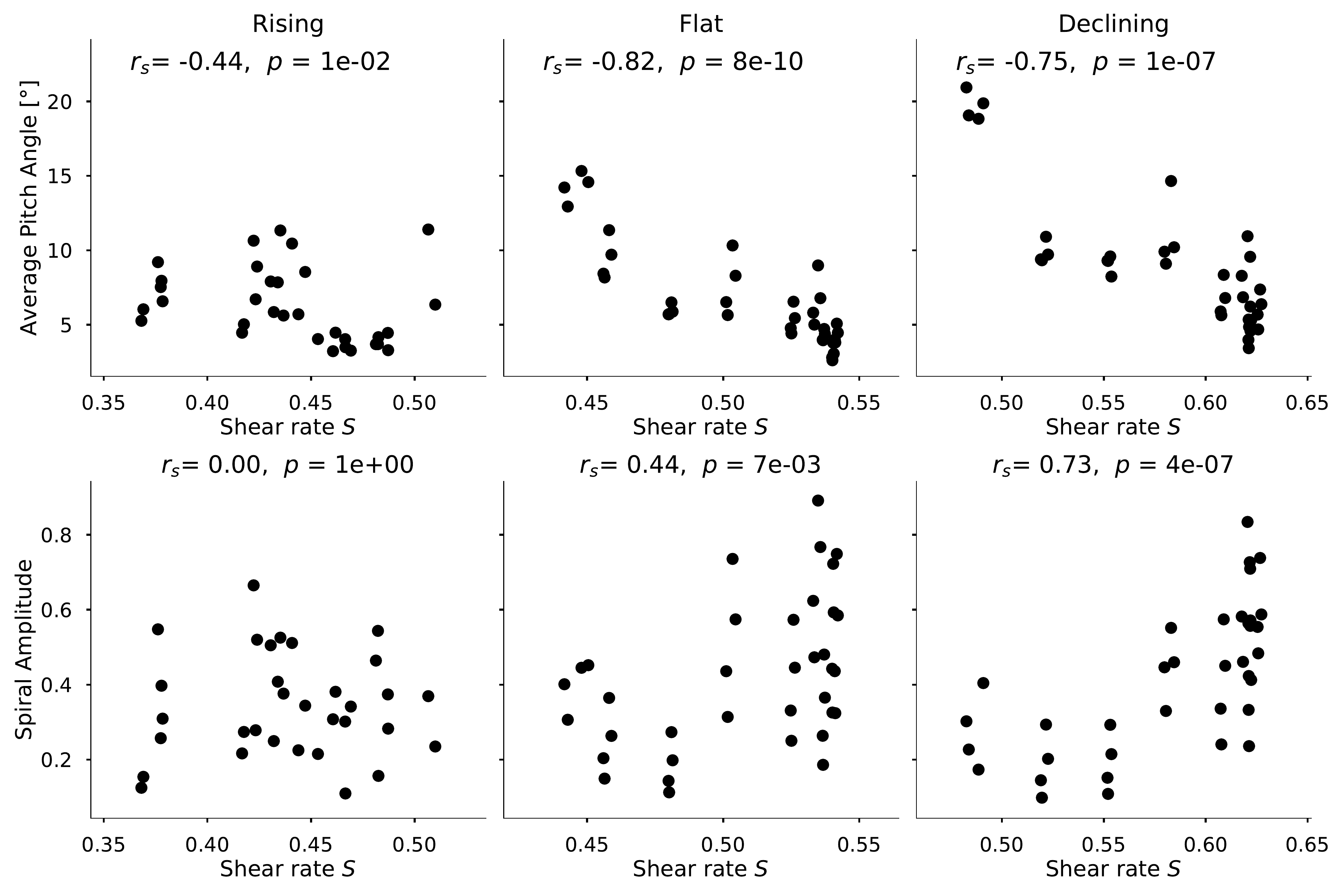}
    \caption{Pitch angle and spiral amplitude vs. Shear rate measured at the OLR in the three galactic models.}
    \label{fig:Shear_correlations}
\end{figure*}

In Figure \ref{fig:Shear_correlations} we show the correlations between the spiral properties and the local shear rate. We found a strong anti-correlation with the pitch angle, but only in the \textit{flat} and \textit{declining} models. The relations with the spiral amplitude are weaker, except on the \textit{declining} models, where we observe a strong correlation.


\section{Influence of the velocity ellipsoid}
\label{sec:velocity_ellipsoid}

Disc heating mechanisms play an important role in the secular evolution of galactic discs, altering the stellar kinematics and increasing the random motion of stars. It has been shown that the shape of the velocity ellipsoid, i.e., the ratio of the vertical and radial velocity dispersion $\sigma_z/\sigma_R$ is strongly correlated with the Hubble type, with late-type galaxies being more anisotropic and early-types being more isotropic \citep{vanderKruit1999, Gerssen2012}. However these results have been questioned by \cite{Pinna2018}, who found a wide range of dispersion ratios from the literature around $\sigma_z/\sigma_R = 0.7 \pm 0.2$. In the solar neighbourhood $\sigma_z / \sigma_R$ ranges from $\sim 0.4$ to $\sim 0.9$ depending on the stellar age, mean orbital radii, or even the dynamical modelling \citep{Mackereth2019, Nitschai2020}.

To study the effects of the velocity ellipsoid on the formation of spiral arms, we re-simulate a subset of 8 galaxies in the \textit{flat} galactic model. We build new initial conditions where the velocity dispersion relation is set to $\sigma_z/\sigma_R = 0.5$ and a more extreme case $\sigma_z/\sigma_R = 0.33$. We chose the subset of simulations to cover a wide range of pitch angles and spiral amplitudes produced by the original simulations with isotropic velocity ellipsoid. 

\begin{figure*}
    \centering
    \begin{subfigure}{0.33 \textwidth}
    \caption*{\Large \hspace{.5cm} $\sigma_z / \sigma_R = 1$}
    \includegraphics[width =  \textwidth]{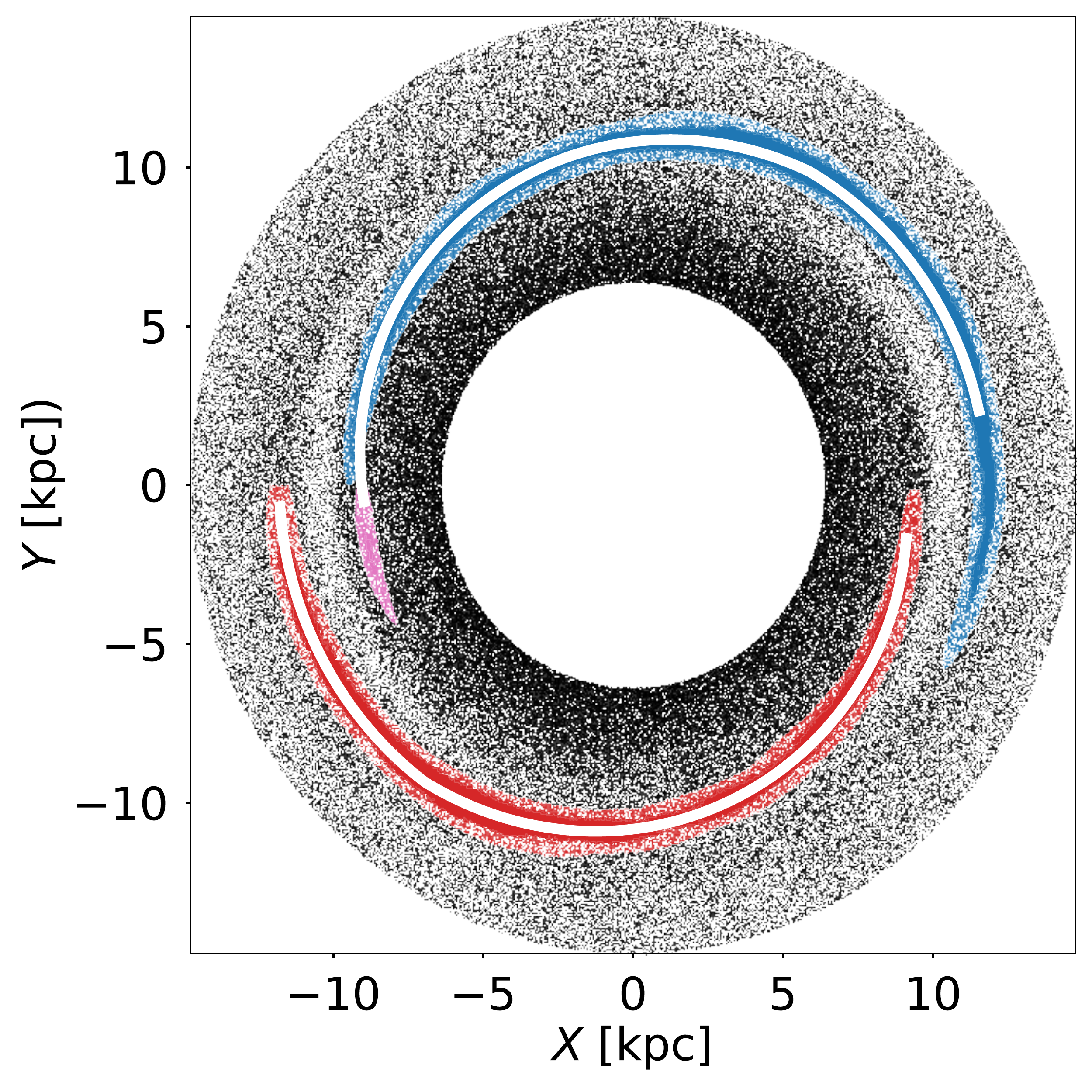}
    \end{subfigure}%
    \begin{subfigure}{0.33 \textwidth}
    \caption*{\Large \hspace{.5cm} $\sigma_z / \sigma_R = 0.5$}
    \includegraphics[width =  \textwidth]{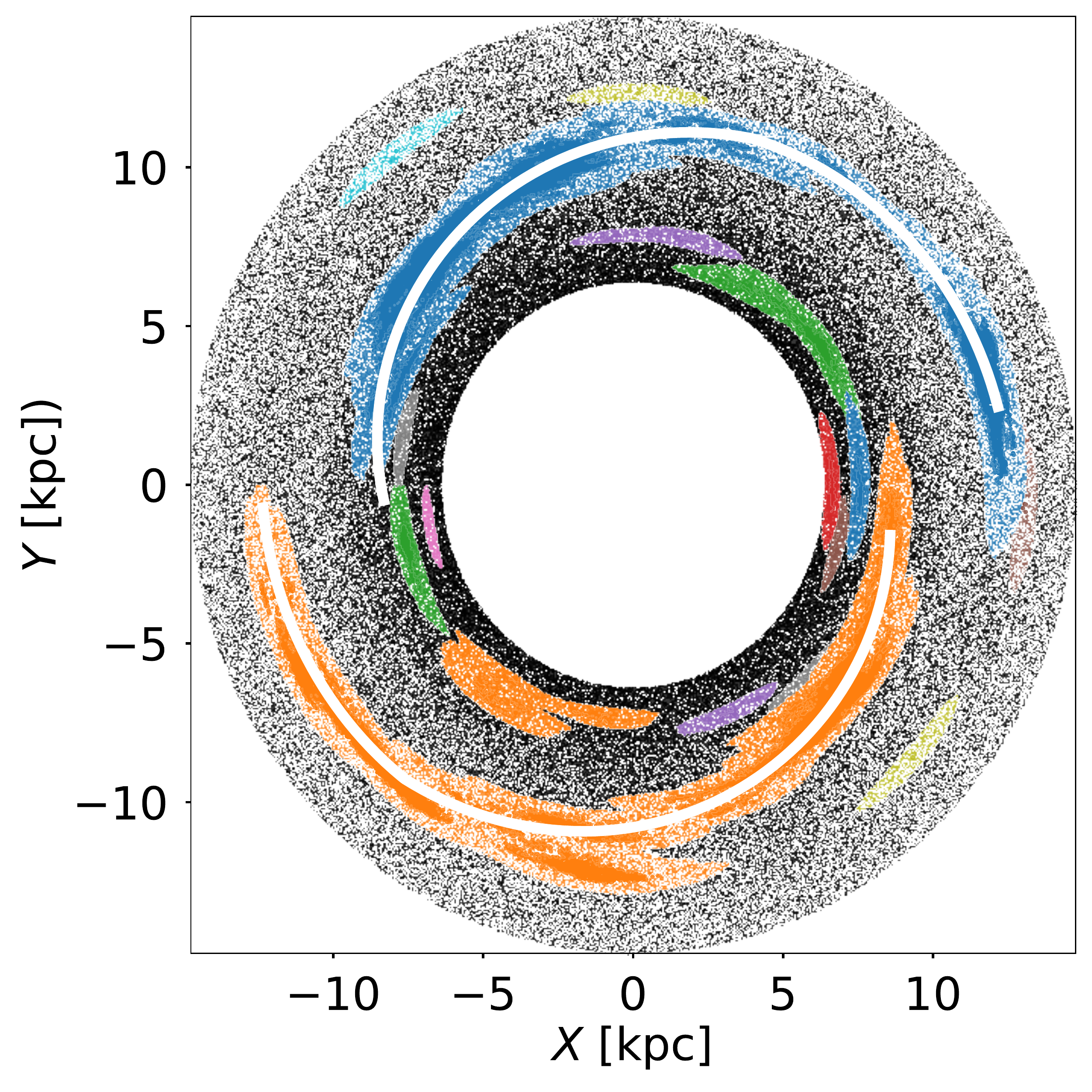}
    \end{subfigure}%
    \begin{subfigure}{0.33 \textwidth}
    \caption*{\Large \hspace{.5cm} $\sigma_z / \sigma_R = 0.33$}
    \includegraphics[width =  \textwidth]{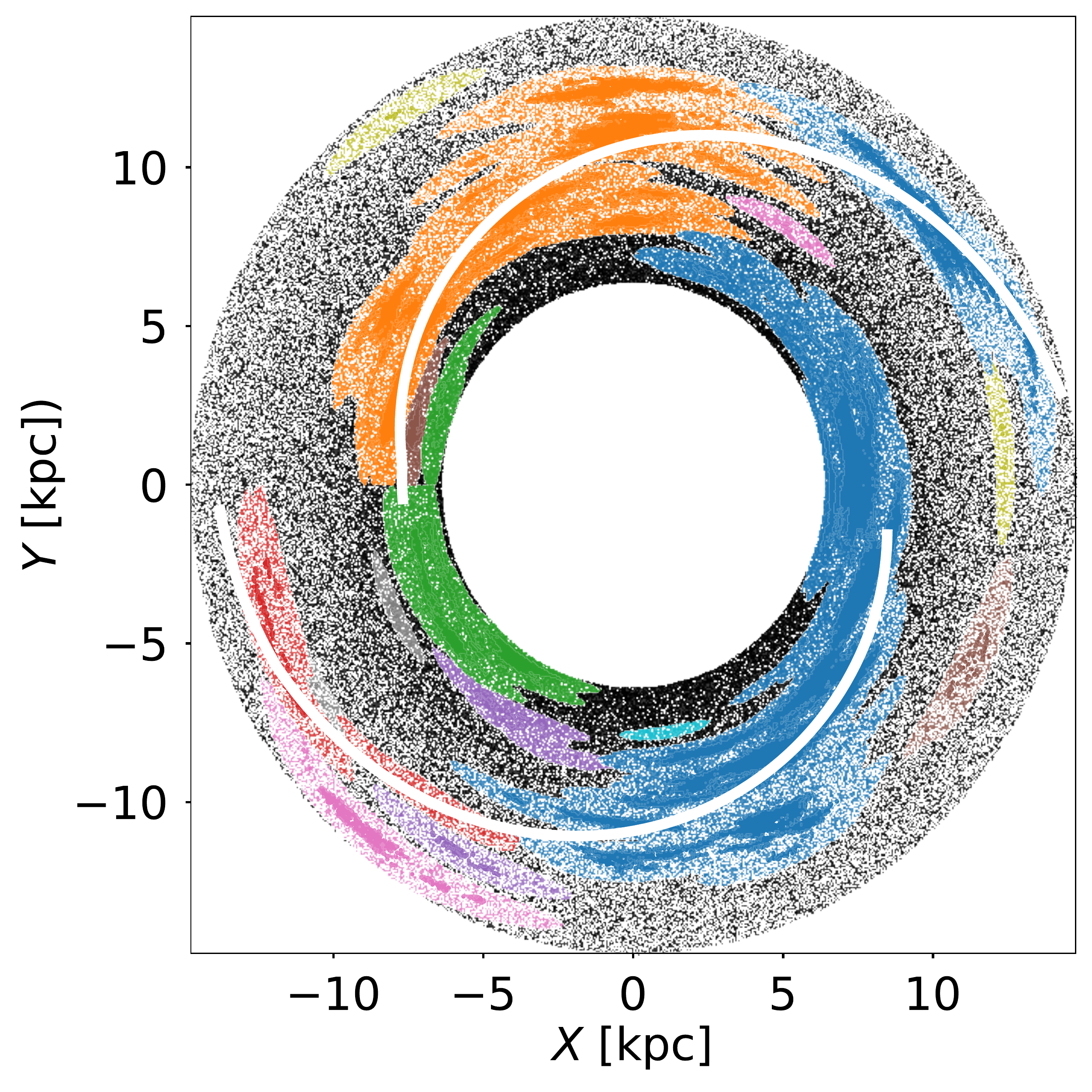}
    \end{subfigure}
    \begin{subfigure}{0.33 \textwidth}
    \includegraphics[width =  \textwidth]{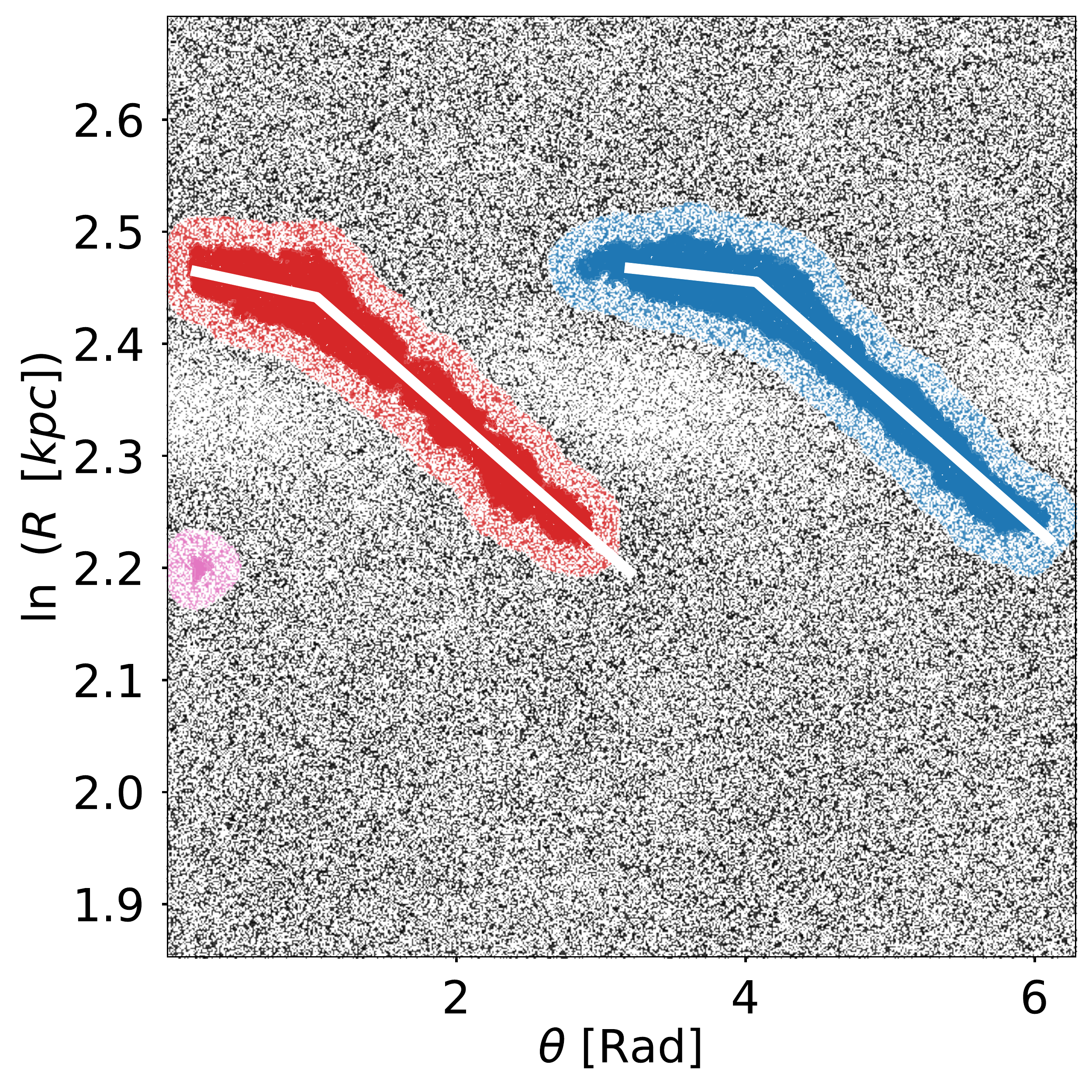}
    \end{subfigure}%
    \begin{subfigure}{0.33 \textwidth}
    \includegraphics[width =  \textwidth]{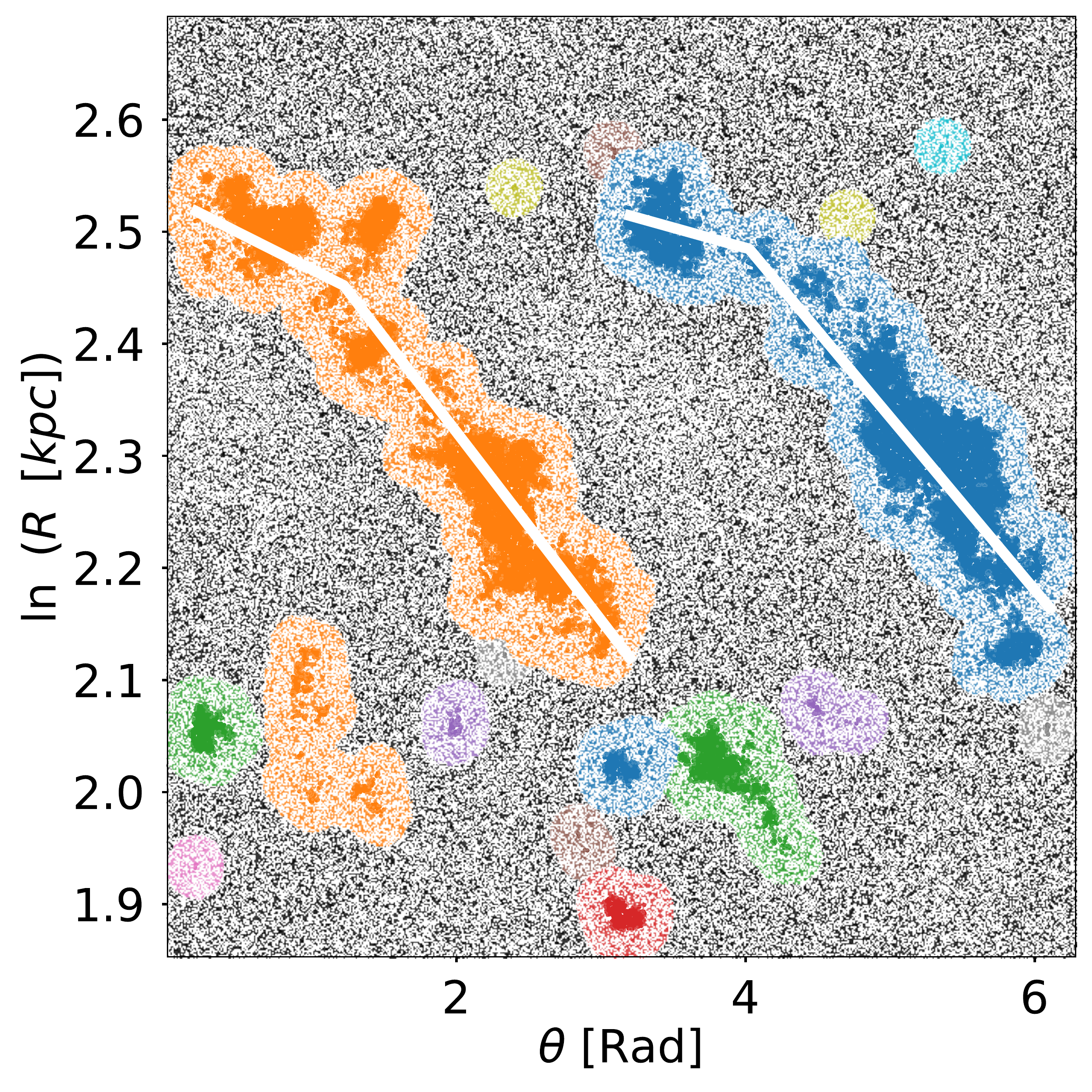}
    \end{subfigure}%
    \begin{subfigure}{0.33 \textwidth}
    \includegraphics[width =  \textwidth]{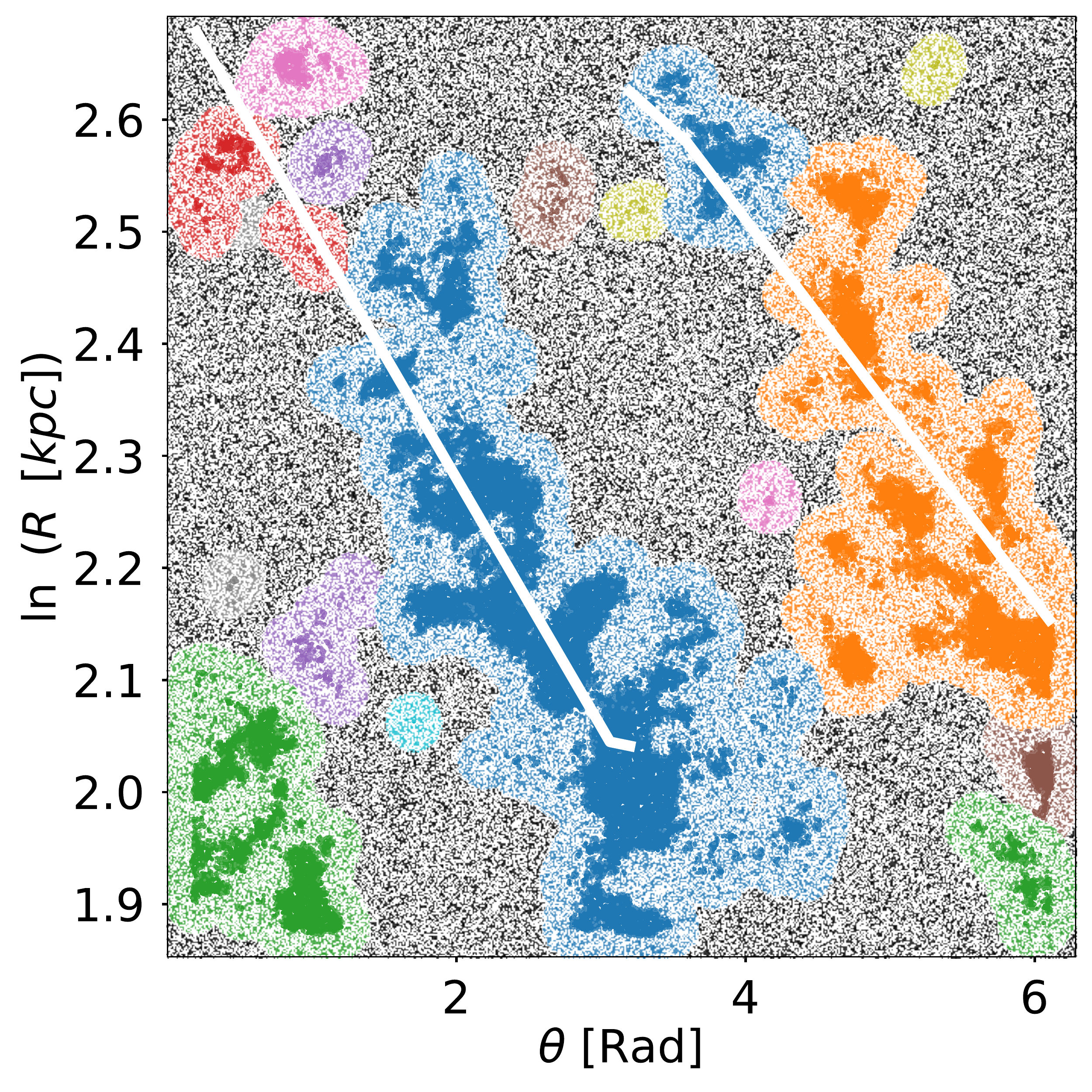}
    \end{subfigure}%
    \caption{Bar induced spiral arms in discs with different velocity ellipsoids. As in Figure \ref{fig:DBSCAN_density}, this is the view of the test particles between $R_{CR}$ and the OLR + 4 kpc at the snapshot where the spiral amplitude is maximum. The top row shows the view in the xy-plane, bottom row shows the \thetalog plane. Test particles classified by \code{DBSCAN} as part of a clusters are coloured. }
    \label{fig:spirals_vel_ellip}
\end{figure*}

The spiral arms produced by the new simulations are still located near the OLR, and peak in amplitude at the same time as the $\sigma_z/\sigma_R = 1$ simulation. In Figure \ref{fig:spirals_vel_ellip} we show a side-by-side comparison of the snapshot we presented in Figure \ref{fig:DBSCAN_density}, but simulated with different velocity ellipsoids. As we decrease the $\sigma_z / \sigma_R$ ratio, the response spirals increase their pitch angle, but result in a less coherent, wider structure with smaller amplitude. 

\begin{figure}
    \centering
    \begin{subfigure}{\linewidth}
    \includegraphics[width =  \linewidth]{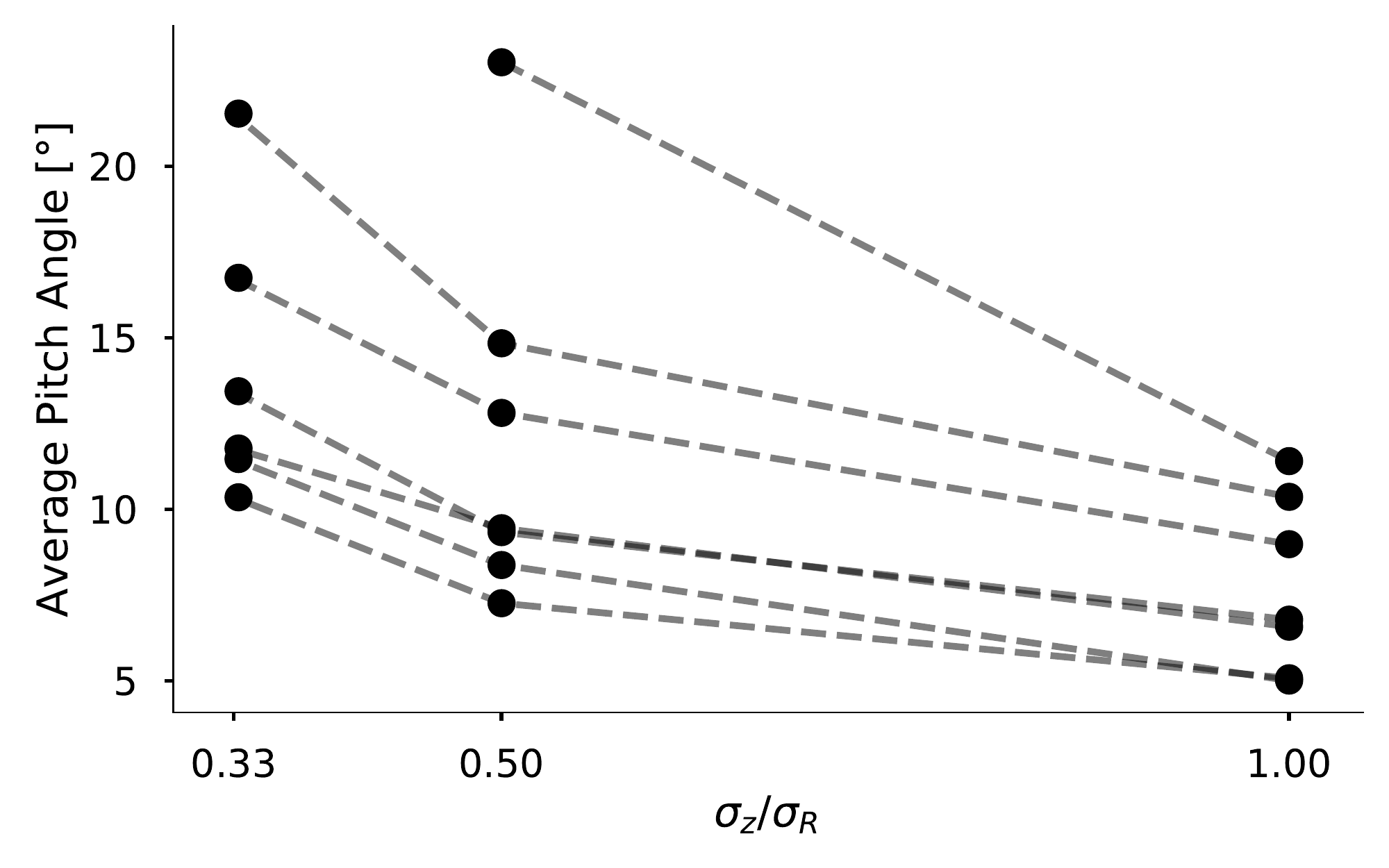}
    \end{subfigure}
    \begin{subfigure}{\linewidth}
    \includegraphics[width =  \linewidth]{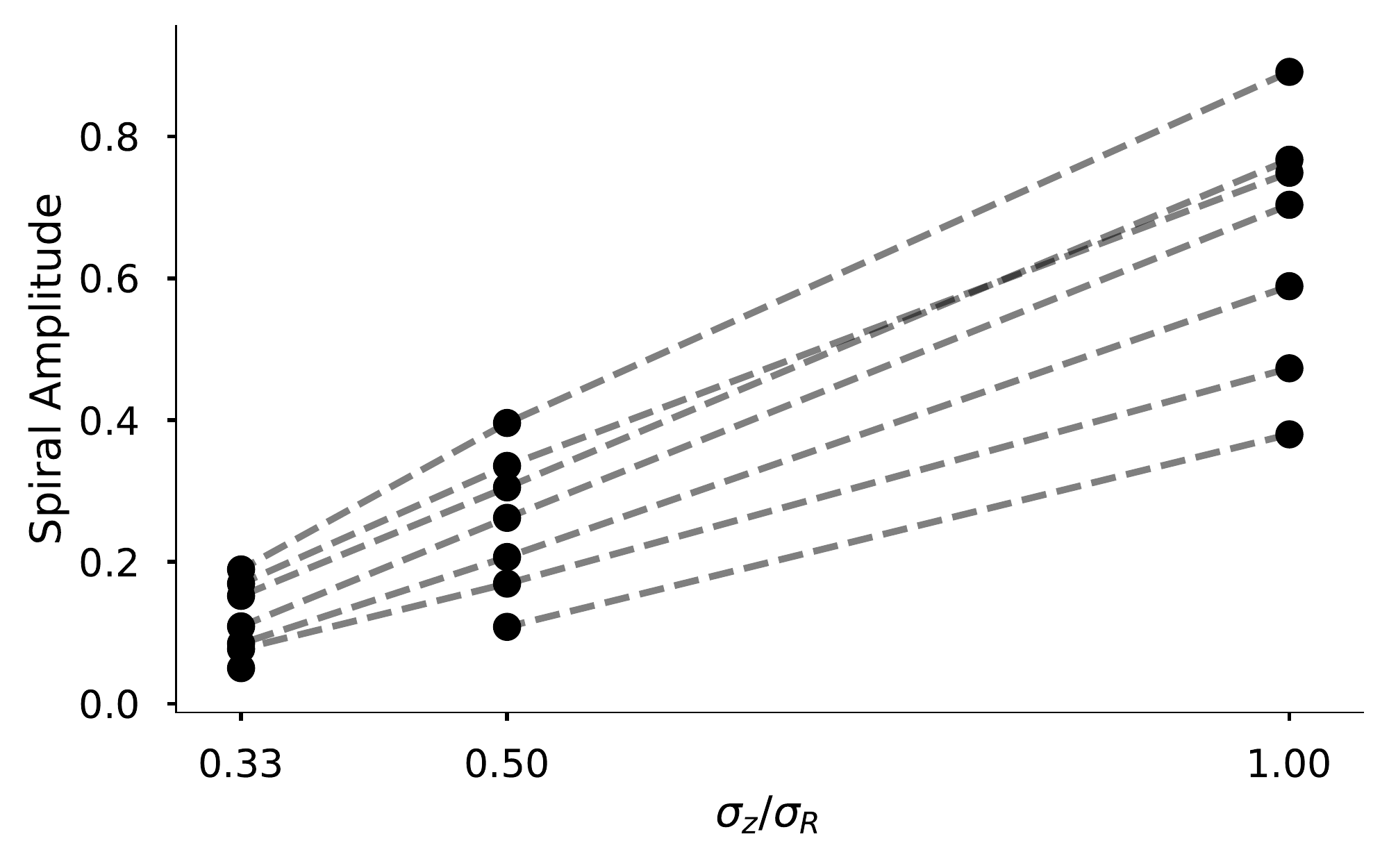}
    \end{subfigure}
    \caption{Pitch angle and spiral amplitude as a function of the velocity dispersion ratio $\sigma_z / \sigma_R$. Simulations that share the same bar model are connected with the grey segmented lines.}
    \label{fig:pitch_amp_vel_ellip}
\end{figure}

In Figure \ref{fig:pitch_amp_vel_ellip} we show the pitch angle and spiral amplitude as a function of the velocity dispersion ratio. We connect with a grey segmented line simulations that share the same bar model. Is clear, from this small subset that the effects the velocity ellipsoid has on the spiral arms are substantial. The more radially heated discs produced very low amplitude spirals. In one simulation, we were not able to detect any spiral structure from the background. 

\section{Bar vs Disc properties}
\label{sec:Bar_vs_Disc}

So far, we have explored how different bar and disc parameters relate with the induced pitch angle and spiral amplitude independently. Using solely their Spearman correlation coefficient and the statistical significance, the parameters that best predict the pitch angle and amplitude are the \Om{} and \QBar{}, respectively. Nonetheless, the differences observed between galactic models suggest the shear or galactic model itself, could also be important features for predicting the spiral properties. Also, some of these parameters are related to each other and cannot be treated as independent variables.

To rank the relative importance of each parameter compared to the rest, we used a random forest regressor \citep{Breiman2001}, trained to predict the spiral properties of our simulations using all bar parameters, the shear rate and the galactic model. The model works by averaging the prediction of multiple decision trees, which try to predict the target variable using a flowchart-like decision, ranking the input variables. 

We set the number of decision trees to 1000 and the maximum number of features used by each tree to 4. This prevents over-fitting, and increases the relative importance of features that are highly correlated. For example, if \Om{} is considered an important variable to predict the pitch angle, a decision tree that does not train with this feature should weight \Rpar{} more heavily, as these two are highly correlated. 

\begin{figure}
    \centering
    \begin{subfigure}{\linewidth} 
    \includegraphics[width=\linewidth]{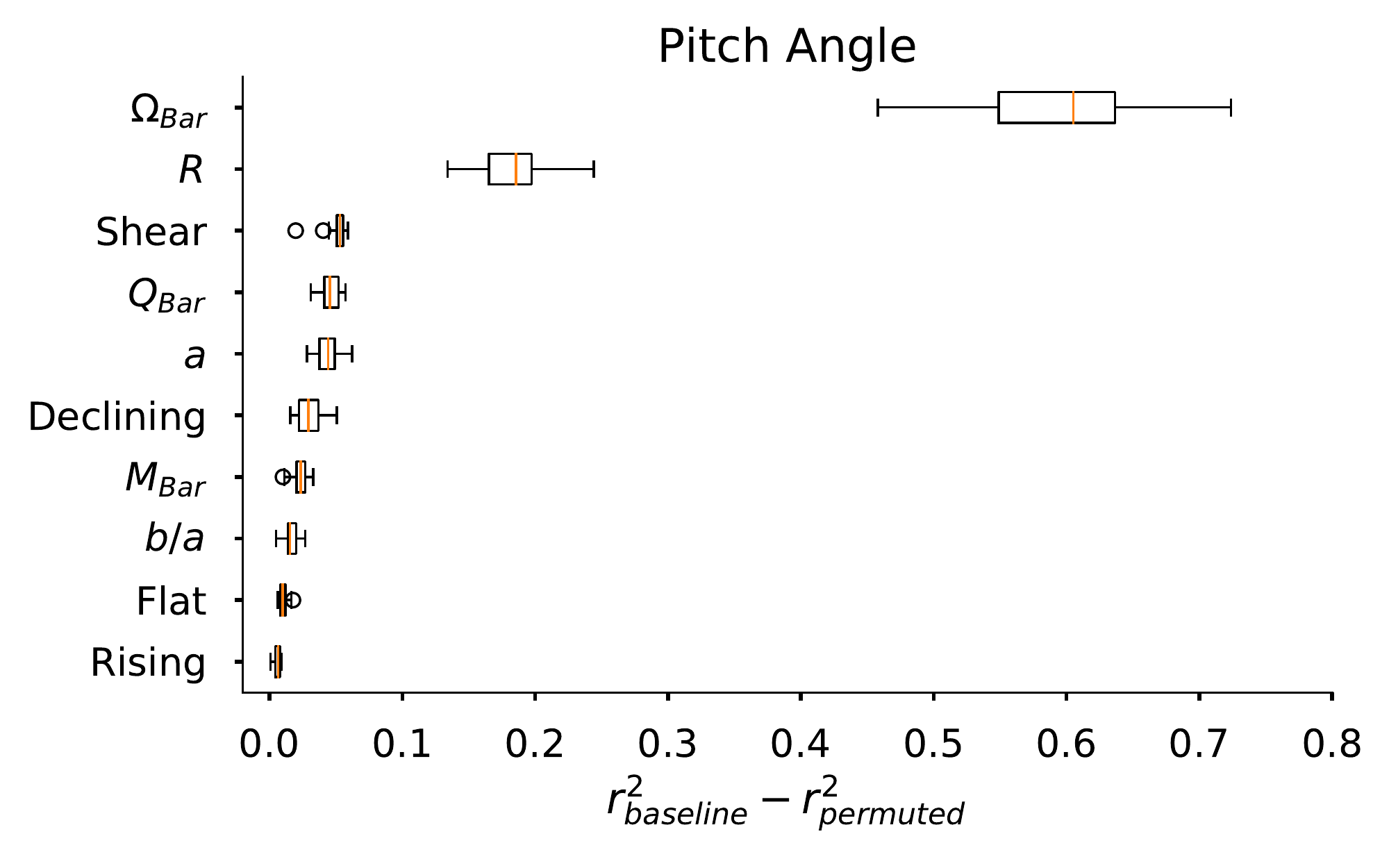}
    \end{subfigure}
    
    \begin{subfigure}{\linewidth} 
    \includegraphics[width=\linewidth]{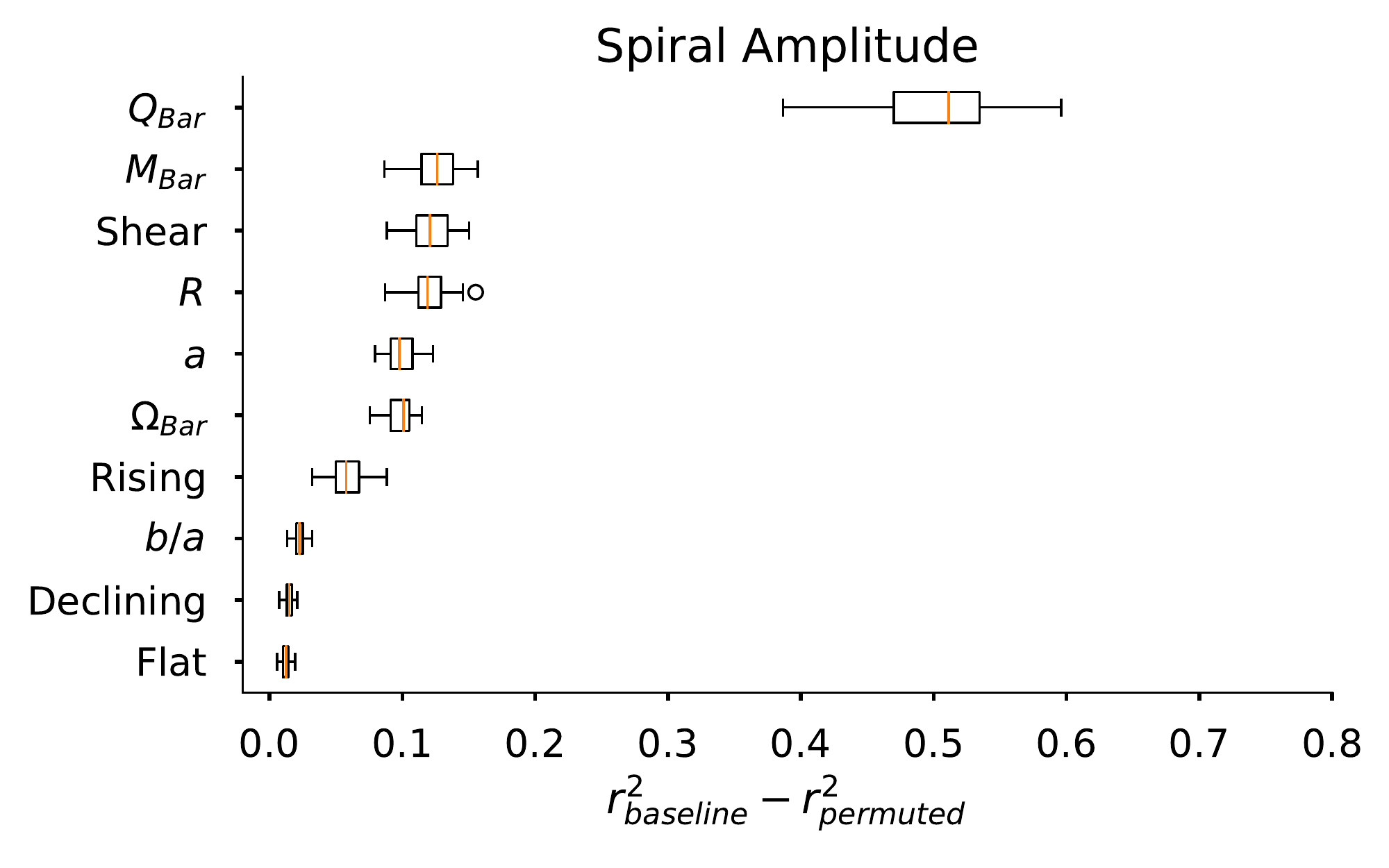}
    \end{subfigure}
    \caption{Permutation feature importance of two random forest regressors trained to predict the pitch angle (top panel) and spiral amplitude (bottom panel). Each feature was permuted 50 times, to produce a distribution of $\Delta r^2$ scores. The orange line shows the median value of the distribution. The box limits are the 25 and 75 percentiles. The lines extend to the maximum and minimum values. Outliers of the distribution are shown with open circles.}
    \label{fig:Importance}
\end{figure}

As a proof of concept, we trained several models using an 80/20 train-test split which consistently achieved a coefficient of determination $r^2 > 0.80$ for both the pitch angle and the spiral amplitude. Since the purpose of these models is to get the relative importance of each feature, we re-train the models using all our simulation data. We used the permutation feature importance technique; where the relevance of each feature is estimated by the difference between the $r^2$ score of the original data ($r^2_{baseline}$) and the score after randomly shuffling such feature ($r^2_{permuted}$) \citep{Breiman2001}. In Figure \ref{fig:Importance} we show the feature importance estimated after shuffling each feature 50 times.

The most important features for predicting the pitch angle are related to the bar rotation rate. This was expected, as \Om{} and \Rpar{} showed the strongest correlations with the pitch angle. In comparison all other parameters are not consider as important, as these only change the $r^2$ score of the prediction by 0.1 or less. This shows that the relative strong correlation with the bar length was actually a consequence of the much more stronger relations with the bar frequency. Notice in particular, that the ''Declining" feature is considered more important than the Flat and Rising features. This is probably due the declining simulations having consistently higher pitch angles as we showed in Figure \ref{fig:Comparing_pitch}.

For the spiral amplitude the most important features are related to the bar strength. Again, this was expected from the strong correlations of \QBar{}, $M_{Bar}$  and $a$. In contrast with the pitch angle, the amplitude model considers the bar frequency (\Rpar{} and \Om{}) and the disc rotation curve ($S$ and Rising) to be as important as the bar mass and length. This could be because of the \textit{rising} simulations forming consistently spirals with lower amplitudes as we showed in Figure \ref{fig:Comparing_amplitudes}. 

Our random forest models do not take into account the velocity dispersion ratio of the stellar disc. However, it is clear from the simulations we studied in Section \ref{sec:velocity_ellipsoid} that the dynamical temperature of the disc is an important feature to consider when predicting the spiral properties. Perhaps equally or more important than the perturber properties or the shape of the velocity curve. How these effects complement or affect the relations discussed, will be explored in a future work.

\section{Discussion and Conclusions}
\label{sec:discussion}

Since our simulations lack the effects of self-gravity, our results cannot predict how the system would evolve after the initial perturbation. Our spirals have typical lifetimes between 2 or 3 dynamical times at the OLR after the bar formation. However, we point out that self-gravity and star formation could increase and highlight the spiral amplitude.
It is expected that these spirals would eventually disappear and fragment into smaller substructures as several N-body simulations have shown. 

The evolution of the pitch angle is unclear. In the Lin-Shu density wave picture, the pitch angle remains constant in time and depends on the global galaxy properties. Several N-body simulations have shown that the pitch angle decreases with time \citep{Grand2013, Pettitt2018}, while others have shown it remains roughly constant, independently of their origin \citep{Mata2019}. Nonetheless, \citet{Mata2019} also observed an increase in the pitch angle after the buckling of the bar, probably because of the mass redistribution. Recently, \cite{Pringle2019} proposed the pitch angle evolves as a decreasing function of time, from an initial measurable maximum $\alpha_{max}$ to a minimum $\alpha_{min}$, as evidenced from the uniform distribution of $\cot \alpha$ of galaxies observed by \cite{Yu2018}. Our results suggest that, when the spirals are induced by a bar perturber, the initial pitch angle is not random nor has a fixed maximum for all galaxies. It mainly depends on the bar pattern speed.

In the majority of our simulations the spirals are tightly wound. In order to produce wide open spirals in barred galaxies like NGC 1365, NGC 1672 and NGC 2903 other interaction mechanisms should be considered. For example, the different nature of spiral arms has also been explained by invariant manifolds \citep{Athanassoula2012, Romero2015}, galaxy encounters \citep{Pettitt2016}, evolution of the mass distribution \citep{Mata2019} or a shearing spiral pattern \citep{Speights2011}.

Although the shear and the rotation curve shape did not appear to be powerful predictors for the spiral properties (in comparison with the bar) they seem to be important on how the disc responds to the perturbations. Our simulations show that bars in galaxies with \textit{rising} rotation curves are less efficient forming a grand design spiral structure. We found evidence in the weaker relationship between the spiral amplitude and \QBar{}, and the 1:1 comparison of the spiral amplitude between galaxy models with same bar parameters. In contrast, bars in galaxies with \textit{declining} rotation curves appear to be the most efficient, having the strongest response with \QBar{}, and consistently producing spirals with higher pitch angles compared with the other two galactic models. 

These results are consistent with observations. Using galaxies from the Spitzer Survey of Stellar Structure in Galaxies ($S^4G$), \cite{Bittner2017} showed that the distribution of flocculent galaxies is statistically different from the multi-arm and grand design galaxies. Flocculent spirals are more common in late-type galaxies, with weaker bars and less concentrated bulges. The low mass concentration suggests this galaxies should have a slowly rising rotation curve. N-body simulations have shown that galaxies with rapidly rising rotation curves form strong flat bars, whereas exponential bars are more typical of slowly rising rotation curves \citep{Combes1993, Athanassoula2002}.

\cite{Diaz2019} measured the pitch angle, spiral strength and bar strength of galaxies in the $S^4G$ survey. They observed the same strong relationship between the spiral and bar strength, even in flocculent galaxies, independently of the method used for the measurement \citep[see also][]{Buta2003, Block2004, Salo2010}. The relationship with flocculent galaxies is consistently weaker (i.e. bars in flocculent galaxies are associated with weaker spirals). They also observed a positive weak correlation between the spiral strength and pitch angle (see their Figs. 17 and E.1).  However, the dispersion in the pitch angle measurements is quite large. We do not find any correlation between the spiral properties independently of the galaxy model used ($r_S=0.01$, $p=0.88$ for all our simulations). Nonetheless, this could be an effect of the spiral arm acting on itself via self-gravity that we cannot capture in our simulations. 

It is possible that our major results could be generalised to other kinds of spiral-producing perturbations (galaxy encounters, tri-axial dark matter halos, giant molecular clouds, etc). That is, the frequency of the perturber relates to the spiral pitch angle, and the strength of the perturbation relates to the spiral amplitude \citep[see e.g.][]{Pettitt2016, Pettitt2018}. 

It is clear that the shape of the velocity ellipsoid and the dynamical state of the disc have an important role in predicting the spiral properties and how the disc responds to different gravitational perturbations. Simulations by \cite{Athanassoula1986} showed that bars delay their growth in dynamically hotter discs. In our tests, spirals generated in discs with isotropic ellipsoids had a greater amplitude, but a smaller pitch angle. As we increased the anisotropy with greater radial dispersion, the spirals unwind, but reduced their amplitude. Is possible that an optimal range of anisotropy is required for discs to be able to form grand design spirals. How the velocity ellipsoid affects the relations we discussed with the bar and the rotation curve remains unclear, and will be explored in a future work.


\section*{Data availability}

The data that support the findings of this study are available from the corresponding author, upon reasonable request.

\section*{Acknowledgements}

We thank the referee C. Struck for his useful comments that improved this paper. We acknowledge DGTIC-UNAM for providing HPC resources on the Cluster Supercomputer Miztli. LGO and LMM acknowledge support from PAPIIT IA101520 grant. LGO acknowledge support from CONACyT scholarship. 




\bibliographystyle{mnras}
\bibliography{references}









\bsp	
\label{lastpage}
\end{document}